# Quantum Entanglement Degree, Mean Positronium Lifetime, and the 3γ/2γ Annihilation-Rate Ratio as Novel PET Biomarkers for Hypoxia – Concept, Challenges, and Predictions

**Bio-Algorithms and Med-Systems – Template for Authors**

# Quantum Entanglement, Positronium Lifetime, and Annihilation Rates as Hypoxia Biomarkers

Paweł Moskal[1,2,3]

[1]Faculty of Physics, Astronomy and Applied Computer Science, Jagiellonian University, Krakow, Poland

[2]Total-Body Jagiellonian-PET Laboratory, Jagiellonian University, Krakow, Poland

[3]Center for Theranostics, Jagiellonian University, Krakow, Poland

**Objective:** Hypoxia is a critical factor in tumor aggressiveness, metastasis, and treatment resistance. Despite its clinical significance, a non-invasive, high-precision method for assessing and mapping tissue oxygenation in vivo remains a major challenge in medicine.

This study explores the potential of positronium imaging and quantum entanglement imaging as next-generation biomarkers for hypoxia. This manuscript introduces a novel method to assess tissue oxygen concentration via the quantum entanglement (QE) of photons originating from positronium—a bound state of an electron and a positron— which is produced within the patient's body during positron emission tomography (PET). We also investigate the possibility of assessing hypoxia by simultaneously detecting positronium lifetime and the positronium decay rate ratio.

**Methods:** We introduce two distinct quantum sensing approaches. *Method 1* utilizes the correlation between oxygen concentration and ortho-positronium (o-Ps) decay rates, relying on the simultaneous measurement of the mean o-Ps lifetime ($\tau_{oPs}$) and the 3γ-to-2γ annihilation rate ratio of ortho-positronium ($R_{oPs-3\gamma/2\gamma}$). *Method 2* introduces a novel hypothesis based on the degree of quantum entanglement (QE) of annihilation photons. This method leverages recent discoveries indicating that the degree of QE is sensitive to the relative contribution of annihilation mechanisms (pick-off vs. conversion), which in turn depends on the oxygen concentration. We estimate the rate of conversion processes as a function of the partial pressure of oxygen ($pO_2$) in water, isopropanol, cyclohexane, isooctane, and adipose tissue. We consider dependence of $R_{oPs-3\gamma/2\gamma}$ and $\tau_{oPs}$, as well as the degree of quantum entanglement, on oxygen pressure in the studied substances. Finally, we derive a formula for $pO_2$ as a function of $R_{oPs-3\gamma/2\gamma}$ and $\tau_{oPs}$ and estimate the measurement accuracy required for these parameters - and for the degree of QE - to sense in-vivo oxygen pressure in the range between hypoxic and physoxic conditions.

**Results:** Theoretical models and quantitative estimates for $R_{oPs-3\gamma/2\gamma}$, $\tau_{oPs}$ and for the degree of quantum entanglement ($C_{QE}$ and $R_{QE}$) as a function of $pO_2$ are provided for water, organic solvents (isopropanol, cyclohexane, isooctane), and adipose tissue. Applying the formulas derived under the working hypothesis that in pick-off process the photons are not entangled, we estimated that for $pO_2 = 0$, the degree of quantum entanglement $C_{QE}$ is equal to 0.890 for adipose, 0.886 for isopropanol, 0.867 for water, 0.818 for cyclohexane, and 0.784 for isooctane. These results indicate that distinguishing between various tissue types requires a precision of $\sigma(C_{QE}) \approx 0.01$. We also estimated the values of $\tau_{oPs}$, $R_{oPs-3\gamma/2\gamma}$, and $R_{3\gamma/2\gamma}$, as well as the changes in $\tau_{oPs}$, $R_{oPs-3\gamma/2\gamma}$ and $C_{QE}$ between physoxic ($pO_2 \approx 50$ mmHg) and hypoxic ($pO_2 = 10$ mmHg) conditions amount to $\Delta\tau_{oPs} = 5$ ps (water), 48 ps (adipose), 182 ps (isopropanol), 187 ps (cyclohexane), 444 ps (isooctane), $\Delta R_{oPs-3\gamma/2\gamma} = 0.000045$ (water), 0.0004 (adipose), 0.0015 (isopropanol), 0.0015 (cyclohexane), 0.0036 (isooctane), and $\Delta C_{QE} = 0.0003$ (water), 0.0019 (adipose), 0.0054 (isopropanol), 0.0106 (cyclohexane), 0.0243 (isooctane). To distinguish between physoxic and hypoxic conditions *in vivo*, the measurement precision $\sigma(\tau_{oPs})$, $\sigma(R_{oPs-3\gamma/2\gamma})$ and $\sigma(C_{QE})$ must be several times higher than the predicted changes in these values.

**Conclusion:** This work establishes a theoretical and methodological foundation for using the quantum entanglement of photons and the properties of positronium as transformative diagnostic tools for the non-invasive assessment and mapping of tissue oxygenation in the human body. It demonstrates that quantum entanglement can serve not only as a tool for improving PET image quality through noise reduction, but also as a completely new category of biomarker. Quantitative estimations of the effect of oxygen pressure on positronium parameters, as well as the event statistics required for hypoxia assessment, are feasible using the multi-photon total-body J-PET scanner based on plastic scintillators. Furthermore, such assessments will be possible with next-generation, high-sensitivity crystal-based PET scanners, provided they are upgraded to support multi-photon signal acquisition for $\tau_{oPs}$ and 3γ-to-2γ imaging, and double Compton scattering for imaging $C_{QE}$ - the degree of quantum entanglement.

## Brief Description of the Work

This manuscript explores a pioneering approach to non-invasive hypoxia assessment by utilizing positronium imaging and quantum entanglement imaging to map oxygenation in the human body. Recognizing hypoxia as a critical factor in tumor aggressiveness, the author investigates two quantum sensing methods to determine the tissue oxygenation (oxygen partial pressure, $pO_2$). *Method 1* utilizes the correlation between oxygen concentration and ortho-positronium (o-Ps) decay rates, relying on the simultaneous measurement of the mean o-Ps lifetime ($\tau_{oPs}$) and the 3γ-to-2γ annihilation rate ratio of ortho-positronium ($R_{oPs\text{-}3\gamma/2\gamma}$). *Method 2* introduces a novel hypothesis based on the degree of quantum entanglement (QE) of annihilation photons. This method leverages recent discoveries indicating that the degree of QE is sensitive to the relative contribution of annihilation mechanisms (pick-off vs. conversion), which in turn depends on the oxygen concentration. The study provides theoretical models and quantitative estimates for water, adipose tissue, and various organic solvents. The results demonstrate that although high precision is required, current and future total-body PET systems (such as the plastic scintillator-based J-PET scanner or crystal-based scanners upgraded with multi-photon detection and double Compton scattering capabilities) can enable clinical translation. Ultimately, this work establishes the methodological foundation for 'quantum-enhanced' medical diagnostics by demonstrating that quantum entanglement can serve not only as a tool for improving PET image quality through noise reduction, but also as a completely new category of biomarker.

# Main Text

## Introduction

Hypoxia, a deficit in tissue oxygenation, is a major feature of solid tumors and one of the fundamental factors related to the development of an aggressive phenotype, including metastasis and treatment resistance [1-7]. Therefore, a method for noninvasive assessment of hypoxia (in particular in vivo mapping of oxygen pressure in a body) would facilitate the decisions on the proper choice of anti-cancer therapy and it is keenly awaited in medicine. The partial pressure of oxygen in a human body varies from ~152 mmHg in inspired air to 75-100 mmHg in arterial blood [8], decreasing to ~50 mmHg in venous blood by the end of circulation, and finally reaching the median values of 30-57 mmHg in tissue under normal physiological condition (physoxia) [1]. Hypoxia levels (deficit in oxygenation) can be distinguished as mild (< 20 mmHg), moderate (< 10 mmHg), and severe (< 1 mmHg) [3].

Positronium imaging [9-17] and quantum entanglement imaging [9, 18, 19] open new perspectives for the in vivo assessment and mapping of hypoxia [20] by sensing and imaging of positronium properties in living organisms [13, 14, 21]. Positronium imaging can be translated to the clinics with the emerging new generation of PET systems with the capability of simultaneous multi-photon detection [22]. Currently, these include the J-PET scanner [13, 14, 23-26], the Biograph Vision Quadra [17, 27-30] the PennPET Explorer [15], the brain-dedicated VRAIN scanner [31, 32], the Prism-PET scanner [33], and the NeuroEXPLORER (NX) brain PET scanner [34].

In the tissue the dissolved molecular oxygen $O_2$ affects the positronium annihilation via conversion and oxidation processes [35-38]. These processes make the rate of positronium annihilation in the tissue dependent on the concentration of molecular oxygen. Moreover, changes in oxygen concentration induce changes in the relative probability of annihilation occurring via pick-off, direct annihilation, and conversion processes. Pick-off occurs when the positron within a positronium annihilates by "picking off" an electron from the surrounding atoms. Direct annihilation refers to the case where a positron annihilates without the prior formation of positronium. Finally, conversion denotes the process in which ortho-positronium (oPs) converts to para-positronium (pPs), or pPs to oPs, through interaction with paramagnetic molecules. Dependence of the relative rate of these processes on oxygen concentration constitutes the basis for the possible application of positronium as a biomarker of hypoxia [20].

In this article we explore two possibilities for quantum sensing of hypoxia. First method relies on the simultaneous measurements of the mean ortho-positronium lifetime ($\tau_{oPs}$) and the 3γ-to-2γ annihilation rate ratio of ortho-positronium ($R_{oPs\text{-}3\gamma/2\gamma}$). While second method relies on the determination of the distribution of the relative angle between the polarization planes of the annihilation photons via measurement of the Compton scattering of these photons. The amplitude of this distribution is the measure (the quantum entanglement witness) of the degree of quantum entanglement of the annihilation photons [21].

The primary inspiration for exploring the application of the degree of Quantum Entanglement (QE) as a hypoxia biomarker stems from the recent discovery that the degree of QE is not maximal for photons originating from $e^+e^-\to2\gamma$ annihilation in matter [21]. The result is presented in Fig. 1. This observation indicates that the degree of QE of annihilation photons depends on the material. Thus, it is natural to hypothesize that the degree of QE may also differ according to tissue type and the degree of hypoxia.  Observation published in [21] suggests that the degree of QE of photons originating from the pick-off process is lower than those from the conversion and other processes. If this is the case, then the degree of QE will depend on the oxygen concentration in the tissue since the relative rate of conversion and pick-off processes changes as a function of the degree of oxygen concentration and hence the degree of hypoxia. The rationale beyond the hypothesis that in case of pick-off annihilation photons are not maximally entangled lies in the fact that the objects are quantum entangled if they originate from the pure quantum state (a pure quantum state is a state about which we know everything that can be known). For example, if the photons originate from a para-positronium in the known quantum state (e.g. in the ground state and in vacuum) then these photons are expected to be maximally quantum entangled. Contrary, in case of the pick-off process, the annihilation occurs via a mixture of different pure states with electron and positron having different relative angular momenta with respect to each other. Therefore, photons originating from such a mixed state will not be maximally entangled.

As oxygen concentration rises, the rate of the conversion process increases while the pick-off rate remains constant. Consequently, the average degree of QE across all annihilations is expected to increase as oxygen concentration grows.

The above discussed method for assessing hypoxia based on the degree of QE is qualitatively different from the method based on $R_{oPs\text{-}3\gamma/2\gamma}$ and $\tau_{oPs}$, since neither $R_{oPs\text{-}3\gamma/2\gamma}$ nor $\tau_{oPs}$ involves quantum entanglement of photons. $R_{oPs\text{-}3\gamma/2\gamma}$ and $\tau_{oPs}$ were proposed each as independent diagnostic parameter [11, 42-44].  Though $R_{oPs\text{-}3\gamma/2\gamma}$ and $\tau_{oPs}$ change as the oxygen concentration changes, their clinical application for hypoxia assessment faces challenges [45]. As will be discussed in detail in this article, e.g.  ex-vivo studies of adipose tissue and cardiac myxoma tumors [46], show that the variation of $\tau_{oPs}$ - and therefore $R_{oPs\text{-}3\gamma/2\gamma}$ - from patient to patient is larger than the changes of these parameters expected due to oxygen concentration difference between hypoxic and physoxic conditions [20].  Therefore, measuring only $\tau_{oPs}$ or only $R_{oPs\text{-}3\gamma/2\gamma}$ will make it difficult to infer the degree of hypoxia based on the absolute $\tau_{oPs}$ or $R_{oPs\text{-}3\gamma/2\gamma}$ values. The consideration of relative values of these parameters with respect to the mean values in the organ or entire patient will be required [44]. In this manuscript we will argue that the simultaneous measurement of $\tau_{oPs}$ and $R_{oPs\text{-}3\gamma/2\gamma}$ can enable the assessment of the degree of hypoxia with a much-reduced bias due to patient-to-patient variations. We elaborate on this method and discuss its model dependence and limitations.

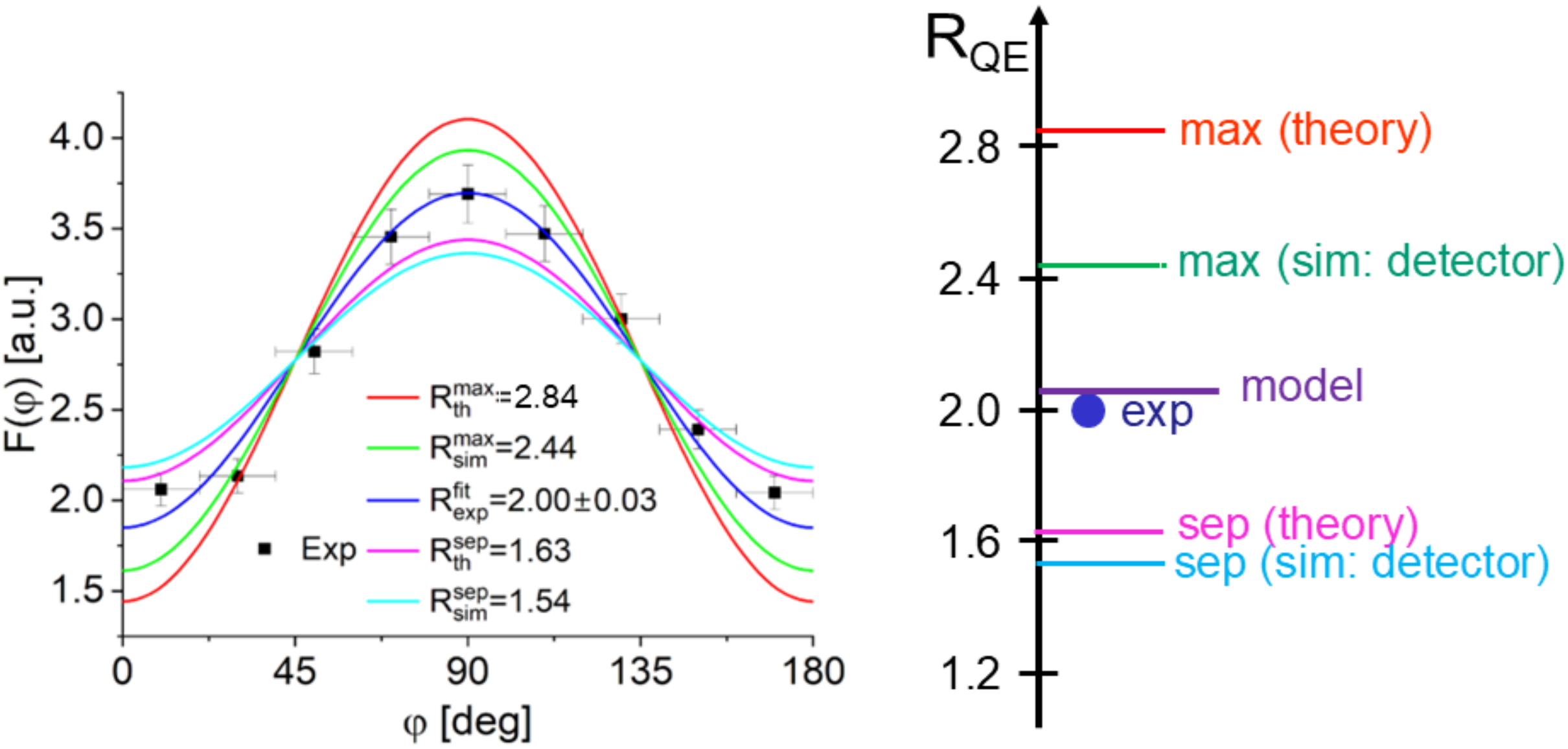

**Fig. 1.** Experimental φ-distribution (black squares) [21] determined using 192-strip J-PET scanner [39, 40]. Blue curve indicates result of the fit of theoretical formula to the experimental data. The detailed explanation of the formula and the degree of entanglement $R_{QE}$ is described further on in section "Basics about quantum entanglement of annihilation photons". The obtained degree of entanglement $R_{QE}$ is equal to Rexp-fit = 2.00 ± 0.03 (blue dot at the right axis), and it is less than expected for the maximally entangled photons and larger than expected for the separable photons [21]. Theoretical distribution for maximal entanglement of Rth-max = 2.84 is shown with red curve. After considering the experimental conditions, the value of Rsim-max = 2.44 is expected (green curve). Magenta curve indicates theoretical expectation for separable photons with Rth-sep = 1.63. After considering the experimental conditions (blue curve), the value of Rsim-sep = 1.54 is expected. The experiment was conducted for positron annihilations in the porous polymer XAD4 in which there was no conversion process because the air was pumped out [21], and pick-off process constituted about 31% of all 2γ annihilations [41]. Horizontal violet line at right axis (model) shows the value of $R_{QE}$ calculated assuming that photons originating from pick-off process are not quantum entangled, while the photons from other annihilation mechanisms (e.g. direct annihilation, annihilation via para-positronium, annihilation via oPs conversion to pPs) are maximally quantum entangled.

In the following sections, we start with a description of the basics of positronium decays in tissue and derive the formulas for $\tau_{oPs}$ and $R_{oPs\text{-}3\gamma/2\gamma}$ as a function of the decay rate due to positronium interaction with oxygen molecules. Next, we introduce the basics about quantum entanglement of annihilation photons and discuss the tissue heterogeneity as the main challenge for detecting hypoxia with positronium. Furthermore, we derive formulas: (i) for the rate of conversion process as a function of the partial oxygen pressure $pO_2$, (ii) for oPs lifetime as a function of the partial oxygen pressure $pO_2$, and (iii) for the ratio of 2γ pick-off annihilations to all 2γ annihilations as a function of the partial oxygen pressure. Next, we present the reasoning that the simultaneous measurement of $\tau_{oPs}$ and $R_{oPs\text{-}3\gamma/2\gamma}$ can enable the assessment of the degree of hypoxia, we elaborate on this method, derive the formula for $pO_2$ as a function of $\tau_{oPs}$ and $R_{oPs\text{-}3\gamma/2\gamma}$, and discuss its model dependence and limitations. Finally, we present in detail the method of sensing hypoxia via the measurement of the degree of quantum entanglement of annihilation photons and predict the dependence of the value of QE witnesses as a function of oxygen concentration for water, for organic liquids as isopropanol, cyclohexane, isooctane and for the adipose tissue.

## Basics about positronium decays in tissue

In the body (Fig. 2A), a positron emitted from the isotope attached to the biomolecule may annihilate with an electron either directly or via formation of positronium [38]. In the tissue positronium intermediates the positron-electron annihilation in about 40% of cases [11, 57, 58]. Positronium may be formed as a long lived (142 ns) spin-one ortho-positronium (oPs), or as a short lived (125 ps) spin-zero para-positronium (pPs). In vacuum oPs decays into 3-photons (oPs → 3γ) and pPs into 2-photons (pPs → 2γ) [38]. In matter (Fig. 2B) the oPs lifetime ($\tau_{oPs}$) is significantly shortened since in addition to self-annihilation into 3-photons, the positron from positronium may pick-off electrons from the surrounding molecular environment (pick-off process) and annihilate predominantly into 2-photons (oPs + M → $e^-$ + 2γ + $M^+$), and 378 times less frequently also to 3-photons (oPs + M → $e^-$ + 3γ + $M^+$) [38]. Positronium in the tissue may also take part in chemical reactions with radiolytic products (e.g. $H_3 0^+$, OH-radicals, hydrated electrons) and with other dissolved substances [37]. In particular, the oPs lifetime may also be shortened by interacting with the dissolved oxygen molecules. oPs may react with a paramagnetic $O_2$ molecule *via* spin-exchange or oxidation processes. In the spin-exchange process, oPs is converted to pPs (oPs + $O_2$ → pPs + $O_2$ → 2γ + $O_2$), which then rapidly decays into two photons, here pPs due to the charge conjugation symmetry cannot decay to 3γ [38]. The oPs oxidation (oPs + $O_2$ → $e^+$ + $O_2^-$) is followed by the direct positron-electron annihilation into 2γ or 3γ. The ratio of cross sections for the annihilation of a free electron-positron pair with small relative velocity into 3γ and into 2γ is in the leading order equal to [59, 60]:

$$\frac{\sigma_{\{3\gamma\}}}{\sigma_{\{2\gamma\}}} = \left(\frac{4}{3\pi}\right)(\pi^2 - 9)\alpha \approx \frac{1}{371.3}. \qquad (0)$$

The next to leading order corrections contribute about 2%. The difference encountered in the prior literature of 1/370 [59], 1/372 [61, 62] and the value of 1/371.3 used in this manuscript are coming from whatever is the latest value of the fine structure constant α (Steven Bass private communication 2025).

The relative ratios and intensities between various processes of electron-positron annihilations and decay rates of positronium relevant for bio-medical applications of positronium are indicated in Fig. 3. In this figure, $f_{pPs}$ and $f_{oPs}$ denote fraction of positronium formations as para-positronium and ortho-positronium, respectively. Based on the spin statistics the ratio is expected to be $f_{oPs}/f_{pPs}$=3. However, in biological material or polymers it may significantly differ from 3. The direct electron-positron annihilation may occur via 2γ or 3γ with the probability of $\beta_2$ and $\beta_3 = 1 - \beta_2$, respectively, with the ratio $\beta_3 / \beta_2 \sim 1/371.3$ (eq. 0). In biological materials the ratio fd/fPs ranges between ~1/4 and ~3/2 [38]. Total Decay rate of positronium in tissue $\lambda_{Ps\text{-}substance}$ is due to the positronium self-annihilation ($\lambda_{Ps\text{-}self}$), pick-off processes ($\lambda_{pick\text{-}off}$), ortho-para conversion reactions ($\lambda_{conv}$) and other chemical processes as e.g. oxidation ($\lambda_{other}$):

$$\lambda_{Ps\text{-}substance} = \lambda_{Ps\text{-}self} + \lambda_{pick\text{-}off} + \lambda_{conv} + \lambda_{other}.$$

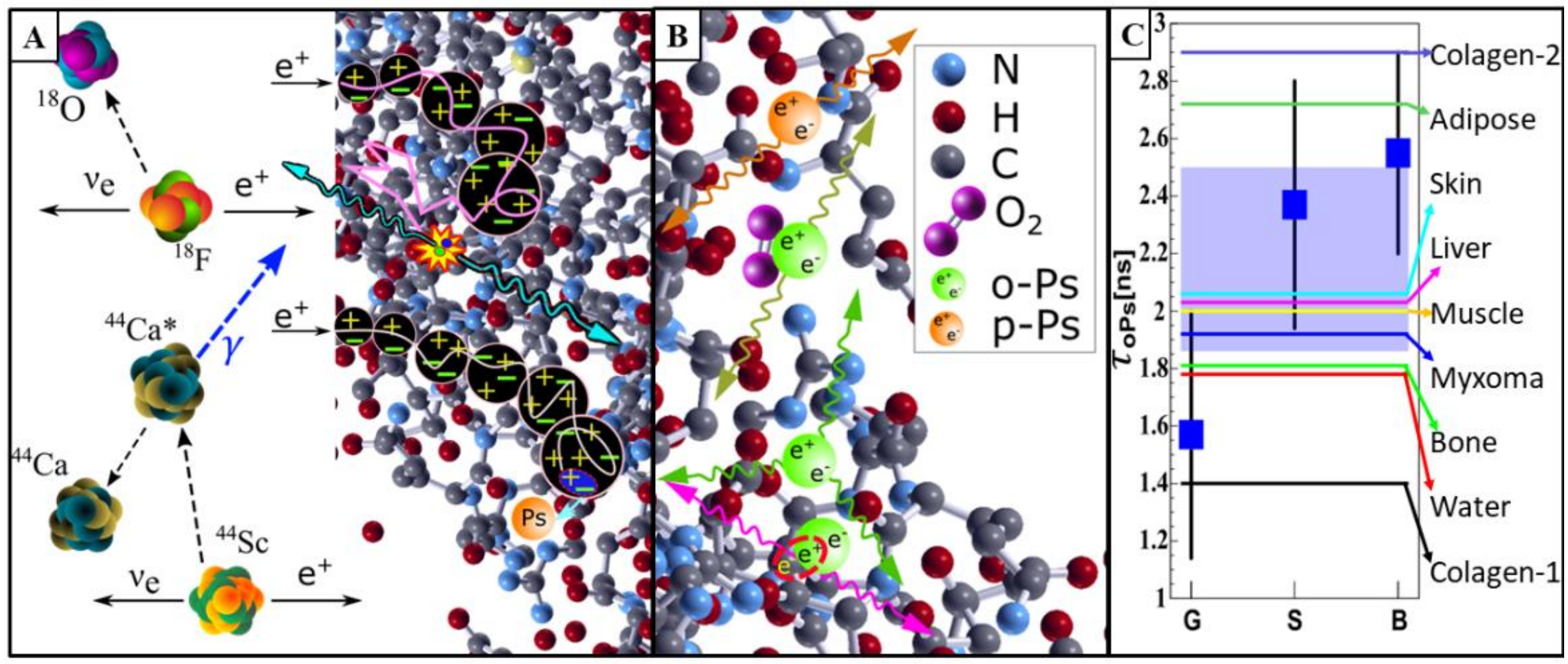

**Fig. 2. (A)** Pictorial illustration of $^{18}$F ($^{18}F \rightarrow {}^{18}O\ e^+ \nu$), and $^{44}$Sc ($^{44}Sc \rightarrow {}^{44}Ca^*\ e^+ \nu \rightarrow {}^{44}Ca\ \gamma_{prompt}\ e^+ \nu$) radionuclides decays. Dashed blue arrow shows the prompt gamma from deexcitation ($^{44}Ca^* \rightarrow {}^{44}Ca\ \gamma_{prompt}$). $^{18}$F isotope is the most frequently used radionuclide in PET [47] and $^{44}$Sc is the most promising radionuclide for positronium imaging [14, 24, 48]. Right part of picture (A) illustrates positrons thermalization at the end of their tracks in the hemoglobin molecule. Ionization sites composed of electrons (-), ions (+) and positron (pink and yellow curves) are shown. In the upper example a thermalized positron scatters in the material and annihilates directly into two photons (blue arrows). In the lower example a quasi-free positronium (yellow dot including +-) is formed in the blob by recombination of thermalized electron with a thermalized positron [49]. Next, quasi-positronium localizes as positronium (orange dot - Ps) in the intra-molecular void. **(B)** Part of the hemoglobin molecule with pictorial representation of possible ways of positronium decays. Para-positronium is marked in orange and ortho-positronium in green. From up to down: (i) pPs → 2γ self-annihilation inside free space between atoms, (ii) oPs annihilation via conversion on oxygen ($oPs + O_2 \rightarrow pPs + O_2 \rightarrow 2\gamma + O_2$), (iii) oPs→3γ self-annihilation inside free space between atoms, (iv) annihilation via interaction with the electron from the surrounding molecule (pick-off effect). The distances and size of atoms are shown to scale with the diameter of positronium twice as large as hydrogen [38]. **(C)** Comparison of oPs mean lifetime for water at 37° [50] and various tissue types: collagen, bones [51], adipose, myxoma [46], skin [52, 53], liver and muscle [54]. Blue shaded area represents data for bio-membranes at 37° in aqueous solution, like DPPC:Cholesterol(60:40) [55] and DPPC:Ceramides(85:15) [56]. The blue squares show results of first *in-vivo* determination of ortho-positronium lifetime in glioblastoma cancer *(G)*, salivary glands *(S)*, and in the healthy brain tissue *(B)* [14].

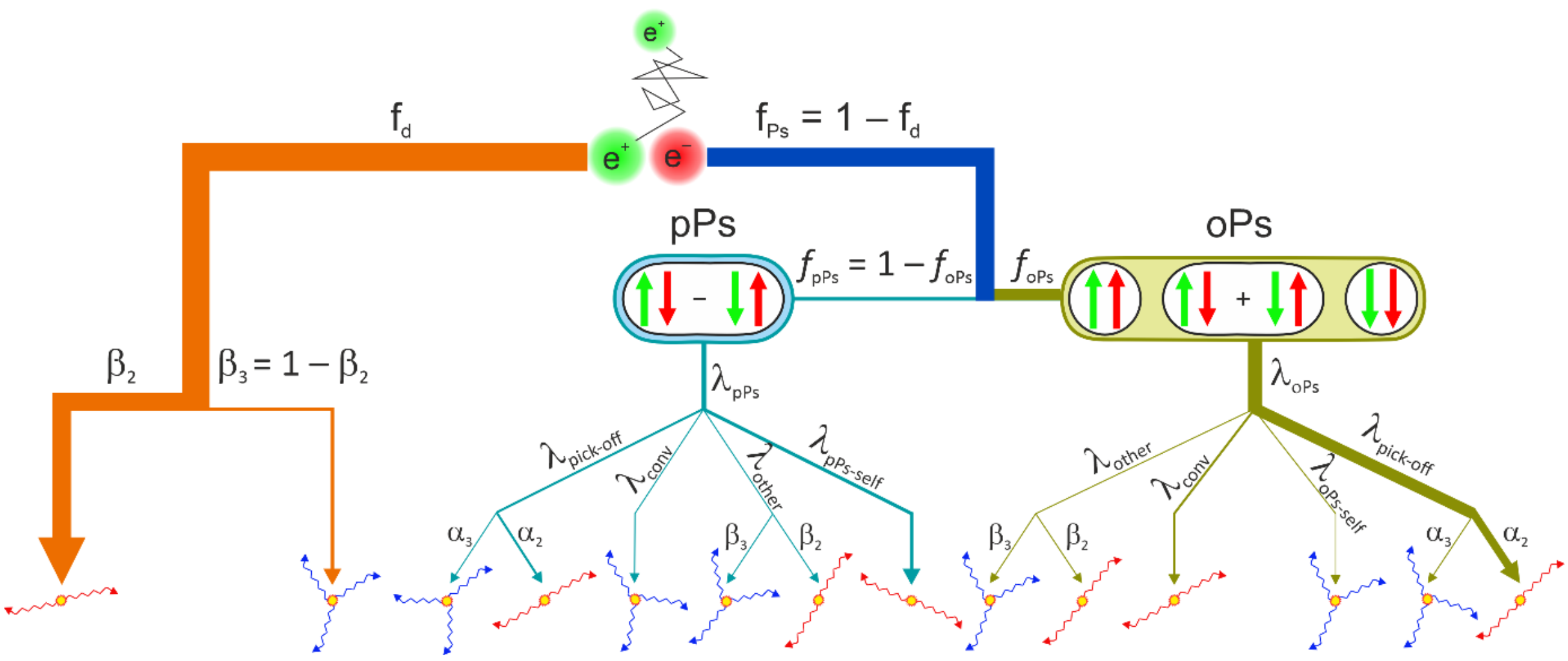

**Fig. 3.** Pictorial representation of the possibilities for electron-positron annihilations in the tissue into 2 and 3 photons. Annihilations into 4 and more photons occur at the level of $10^{-6}$ and are not relevant for bio-medical applications [38]. $f_d$ indicates fraction of direct annihilations and $f_{Ps}$ =1- $f_d$ indicates fraction of annihilations via positronium formation. More details about this figure are available in the text. This scheme is an improved version of similar figure in reference [38].

The self-annihilation rates for para-positronium $\lambda_{p\text{-}Ps\text{-}self}$ = 7990.9 $\mu s^{-1}$ [63] is more than thousand times larger the one for orthopositronium $\lambda_{o\text{-}Ps\text{-}self}$ =7.0401 $\mu s^{-1}$ [64]. $\lambda_{pick\text{-}off}$ depends on the material (Table I) but it is the same for ortho-positronium and for para-positronium [38]. In water with saturated $O_2$ (1400 μmol/L), when the conversion on $O_2$ and oxidation is maximal, the relative strength of decay rates is as follows [20]:

$$\lambda_{pPs\text{-}self}\ (7990.9\ \mu s^{-1}) >> \lambda_{pick\text{-}off}\ (513\ \mu s^{-1}) >> \lambda_{conv} + \lambda_{other}\ (28\ \mu s^{-1}) > \lambda_{oPs\text{-}self}\ (7.0401\ \mu s^{-1}).$$

In case of pick-off process the relative probability of annihilations into 3-photons ($\alpha_3$) and 2-photons ($\alpha_2$ = 1 - $\alpha_3$) is equal to $\alpha_3$ / $\alpha_2$ ~ 1/378 [38]. For other annihilation mechanisms as oxidation, we assume that relative rate into 2 and 3 photons is the same as in case of direct annihilation $\beta_3$ / $\beta_2$. In summary, in this work we assume that $\alpha_2$ = 378/379, $\alpha_3$ = 1/379, $\beta_2$ = 371.3/372.3, $\beta_3$ = 1/372.3.

The mean lifetime of ortho-positronium $\tau_{oPs}$ is an inversed value of the total decay rate of ortho-positronium $\lambda_{oPs}$, and may be expressed as:

$$\tau_{oPs} = \frac{1}{\lambda_{oPs}} = \frac{1}{\lambda_{\text{oPs-self}} + \lambda_{\text{pick-off}} + \lambda_{\text{conv}} + \lambda_{\text{other}}}, \quad (1)$$

and analogously for para-positronium:

$$\tau_{pPs} = \frac{1}{\lambda_{pPs}} = \frac{1}{\lambda_{\text{pPs-self}} + \lambda_{\text{pick-off}} + \lambda_{\text{conv}} + \lambda_{\text{other}}}. \quad (2)$$

Conversion and "other" processes will depend on the gas dissolved in the tissue. In this work we consider the case with the molecular oxygen dissolved in the tissue. In this case oxidation of oPs is the main "other" process in addition to conversion [36, 37]. We assume that other processes constitute a fraction ω of the conversion process, with value of ω ranging between ω = 0.1 to ω= 0.2 [36, 37]. Thus, decay rate due to the interaction of positronium with oxygen $\lambda_{O2}$ reads:

$$\lambda_{\mathrm{O}_2} = \lambda_{\mathrm{O}_2\text{-conv}} + \lambda_{\mathrm{O}_2\text{-other}}\,, \tag{3}$$

where

$$\omega \equiv \frac{\lambda_{\mathrm{O}_2\text{-other}}}{\lambda_{\mathrm{O}_2\text{-conv}}}\,, \tag{4}$$

and hence:

$$\lambda_{\mathrm{O}_2\text{-conv}} = \left(\frac{1}{1+\omega}\right)\lambda_{\mathrm{O}_2} \quad \text{and} \quad \lambda_{\mathrm{O}_2\text{-other}} = \left(\frac{\omega}{1+\omega}\right)\lambda_{\mathrm{O}_2}\,. \tag{5}$$

In this work we will assume that ω= 0.15. Here $\lambda_{\mathrm{O}_2\text{-conv}}$ denotes decay rate of positronium via conversion on oxygen and $\lambda_{\mathrm{O}_2\text{-other}}$ denotes other processes involving interaction of positronium with oxygen.

The overall 3γ to 2γ rate ratio $R_{3\gamma/2\gamma}$ can be expressed as:

$$R_{3\gamma/2\gamma} = \frac{\left[f_d\beta_3 + f_{Ps}f_{pPs}\left(\alpha_3\lambda_{\text{pick-off}} + \lambda_{\mathrm{O}_2\text{-conv}} + \beta_3\lambda_{\mathrm{O}_2\text{-other}}\right)/\lambda_{pPs} + f_{Ps}f_{oPs}\left(\lambda_{\text{oPs-self}} + \alpha_3\lambda_{\text{pick-off}} + \beta_3\lambda_{\mathrm{O}_2\text{-other}}\right)/\lambda_{oPs}\right]}{\left[f_d\beta_2 + f_{Ps}f_{pPs}\left(\alpha_2\lambda_{\text{pick-off}} + \lambda_{\text{pPs-self}} + \beta_2\lambda_{\mathrm{O}_2\text{-other}}\right)/\lambda_{pPs} + f_{Ps}f_{oPs}\left(\alpha_2\lambda_{\text{pick-off}} + \beta_2\lambda_{\mathrm{O}_2\text{-other}} + \lambda_{\mathrm{O}_2\text{-conv}}\right)/\lambda_{oPs}\right]} \tag{6}$$

The above formula is derived for the case when (only) oxygen is dissolved in the substance. It is derived using Fig. 3, which pictorially indicates the most relevant processes leading to 3γ and 2γ annihilations. In general $R_{3\gamma/2\gamma}$ depends on: (i) the size of the free spaces between atoms via $\lambda_{\text{pick-off}}$, (ii) the concentration of paramagnetic molecules (via $\lambda_{O2\text{-conv}}$ and $\lambda_{O2\text{-other}}$ in case of $O_2$ molecules), (iii) the density of voids reflected in the value of $f_{Ps}$, and (iv) the para-positronium to ortho-positronium production ratio $f_{pPs}$ / $f_{oPs}$. It is also important to note that due to the more than one thousand times larger self-annihilation of para-positronium with respect to ortho-positronium, the pickoff and conversion processes are affecting significantly only the mean ortho-positronium lifetime (which changes from 142 ns in vacuum to few ns in tissue) while the mean lifetime of para-positronium changes only by tens of ps [38]. Since ortho-positronium properties are much more affected in the medium than para-positronium properties we introduce also a ratio $R_{oPs\text{-}3\gamma/2\gamma}$ defined as a ratio of oPs decay rate into 3γ to decay rate into 2γ. The formula for $R_{oPs\text{-}3\gamma/2\gamma}$ expressed explicitly in equation (7) reads:

$$R_{\text{oPs-}3\gamma/2\gamma} = \frac{\lambda_{\text{oPs-self}} + \alpha_3\lambda_{\text{pick-off}} + \beta_3\lambda_{\mathrm{O}_2\text{-other}}}{\alpha_2\lambda_{\text{pick-off}} + \beta_2\lambda_{\mathrm{O}_2\text{-other}} + \lambda_{\mathrm{O}_2\text{-conv}}} \tag{7}$$

Formula (7) is much simpler than equation (6) and most importantly $R_{oPs\text{-}3\gamma/2\gamma}$ does not depend on $f_{Ps}$ and on $f_{pPs}$ / $f_{oPs}$ ratio. Therefore, the value of $R_{oPs\text{-}3\gamma/2\gamma}$ is much more sensitive to the variation of oxygen concentrations in tissue than $R_{3\gamma/2\gamma}$, and $R_{oPs\text{-}3\gamma/2\gamma}$ is less sensitive to the tissue heterogeneity than $R_{3\gamma/2\gamma}$.

The above introduced formulas are simplifications for the case if only one void type is present. In general, one would need to consider the distribution of the size of voids in the material and hence the distribution of different $\lambda_{\text{pick-off}}$. Therefore, the above formulae and λ parameters in Fig. 3 may be considered as effective parameters describing the phenomena in a simplified way.

## Basics about quantum entanglement of annihilation photons

Photons are spin-1 particles characterized by their momentum and polarization, with two transverse polarization states for real photons. In the linear polarization basis, the 2γ state |ψ> originating e.g. from pPs can be written as $|\psi\rangle = 1/\sqrt{2}\,(|H\rangle_1 \otimes |V\rangle_2 + |V\rangle_1 \otimes |H\rangle_2)$, where |H> and |V> denote the horizontal and vertical polarized states, and the symbol ⊗ refers to a tensor product. Photons originating from the decay of positronium are expected to be quantum entangled in polarization and exhibit non-local correlations [65]. The entanglement of the 2γ state |ψ> manifests itself in the fact that there is no choice of basis |A>,|B> in which the state could be described by a single tensor product of (|A>⊗|B>) - this we call entanglement [38]. Moreover, Bose-symmetry and parity conservation in the decay of pPs imply that the state of the resulting two photons is maximally entangled and that the photons polarizations are orthogonal to each other, $\boldsymbol{\varepsilon}_1 \perp \boldsymbol{\varepsilon}_2$ [65]. Quantum entanglement of the emitted photons in positronium decays is interesting as a fundamental physics issue [65-74]. It may also have interesting implications for medical diagnostics [18, 19, 21, 75-96].

Photons with energy in MeV range interact with single electrons, therefore their polarization cannot be studied using optical methods. However, polarization direction of such highly energetic photons can be determined by Compton scattering (Fig. 4A). The Compton scattering of photons is most likely in a plane perpendicular to the polarization of the incoming photon [98], and therefore a vector product **(ε = k x k' / | k x k'|)** of momentum vectors of initial **k** and scattered photon **k'** may be a measure of polarization plane orientation of the primary photon at the moment of scattering [79, 86, 99, 100]. For the 2γ annihilation process (Fig. 4A), when each γ interacts via Compton scattering with an electron, one can estimate the angle between the polarization planes of photons $|\eta_1 - \eta_2|$ by measurement of the relative angle φ between the scattering planes. The distribution of φ may serve for studies of quantum entanglement [65]. Fig. 4B shows theoretical distributions of the relative angle φ between polarization orientations of two photons in cases when: (i) the state of the two photons is maximally QE (solid red), (ii) the photons scatter independently of each other (dashed green), and for the case (iii) when the photons' are uncorrelated (dotted black), e.g., when the photons originate from two different pPs decays. Fig. 4B presents results for the scatterings angles $\theta_1 = \theta_2 = 81.7°$, for which the correlation is the highest [65]. The dependence of the correlation on the scattering angle θ shown in Fig. 4C.

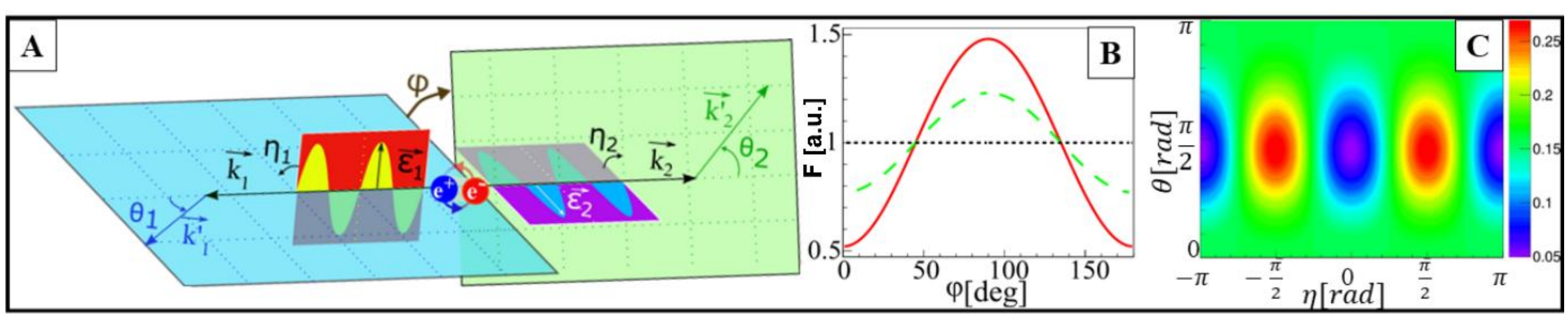

**Fig. 4. (A)** Schematic illustration [38] of Compton scattering of two photons originating from pPs annihilation. Due to the momentum conservation ($\mathbf{k}_1 = -\mathbf{k}_2$), these photons propagate back-to-back along the same axis in the pPs rest frame. $\theta_1$ and $\theta_2$ denote the scattering angles, $\eta_1$ and $\eta_2$ denote the angles between the scattering planes and the polarization planes ($\boldsymbol{\varepsilon}_1$ and $\boldsymbol{\varepsilon}_2$), respectively. φ indicates the relative angle between the scattering planes. **(B)** Distributions of φ for the maximally QE photons (solid red curve), for the case of independent Compton interaction of the two photons (dashed green curve), and for the case when photon polarizations are uncorrelated (dotted black line). The result (green and red curve) is shown for the case of $\theta_1 = \theta_2 = 81.7°$, for which the correlation is the highest. **(C)** Plot of the normalized cross section for γ scattering on an electron as a function of the scattering angle θ and the azimuthal angle η [97]. Around θ = 81.7° the biggest variation is seen.

The φ-distribution depending on the scattering angles $θ_1$ and $θ_2$ is described by the formula [73]:

$$F(\theta_1, \theta_2, C_{QE}, \varphi) = F_0\left(1 - C_{QE}\, A(\theta_1)\, A(\theta_2) \cos(2\varphi)\right) \quad (8)$$

where $C_{QE} A(θ_1) A(θ_2) \equiv V$ is referred to as visibility V, which is describing the interference contrast, and $C_{QE}$ is a measure of photons correlation [73]. Thus, $C_{QE}$ is a measure of the degree of quantum entanglement. For maximally quantum entangled photons $C_{QE}$ = 1, while for photons propagating independently of each other $C_{QE}$ = ½. $A(θ_1)$ and $A(θ_2)$ denote analyzing powers in Compton polarimeter. For annihilation photons from $e^- e^+ \rightarrow 2γ$, with energy of each photon equal to the mass of the electron, A(θ) reads [101]:

$$A(\theta) = \frac{\sin^2\theta\,(2 - \cos\theta)}{2 + (1 - \cos\theta)^3} \quad (9)$$

Experimentally, visibility $V(θ_1,θ_2)$ can be determined from the measured $F(θ_1,θ_2,C_{QE},φ)$ distribution as:

$$V(\theta_1, \theta_2, C_{QE}) = \frac{F(\theta_1, \theta_2, C_{QE}, \varphi = 90^\circ) - F(\theta_1, \theta_2, C_{QE}, \varphi = 0^\circ)}{F(\theta_1, \theta_2, C_{QE}, \varphi = 90^\circ) + F(\theta_1, \theta_2, C_{QE}, \varphi = 0^\circ)}. \quad (10)$$

Coefficient $C_{QE}$ is independent of the scattering angles $θ_1$ and $θ_2$ and therefore it is a convenient measure of the degree of quantum entanglement [73].

The degree of QE may also be measured by parameter $R_{QE}(θ_1,θ_2,C_{QE})$ which is the ratio of the probabilities of the scattering at ϕ = 90° and ϕ = 0° [82]:

$$R_{QE}(\theta_1, \theta_2, C_{QE}) = \frac{F(\theta_1, \theta_2, C_{QE}, \varphi = 90^\circ)}{F(\theta_1, \theta_2, C_{QE}, \varphi = 0^\circ)}. \quad (11)$$

For maximally QE photons, the value of $R_{QE}$ reaches $R_{QE}max$ = 2.84 at angles $θ_1 = θ_2$ = 81.7°, while for independent (separable) photons the value of $R_{QE}$ for these angles is equal to $R_{QE}max\text{-}sep$ = 1.63. The dependence of $R_{QE}$ as a function of $θ = θ_1 = θ_2$ is shown in Fig. 5.

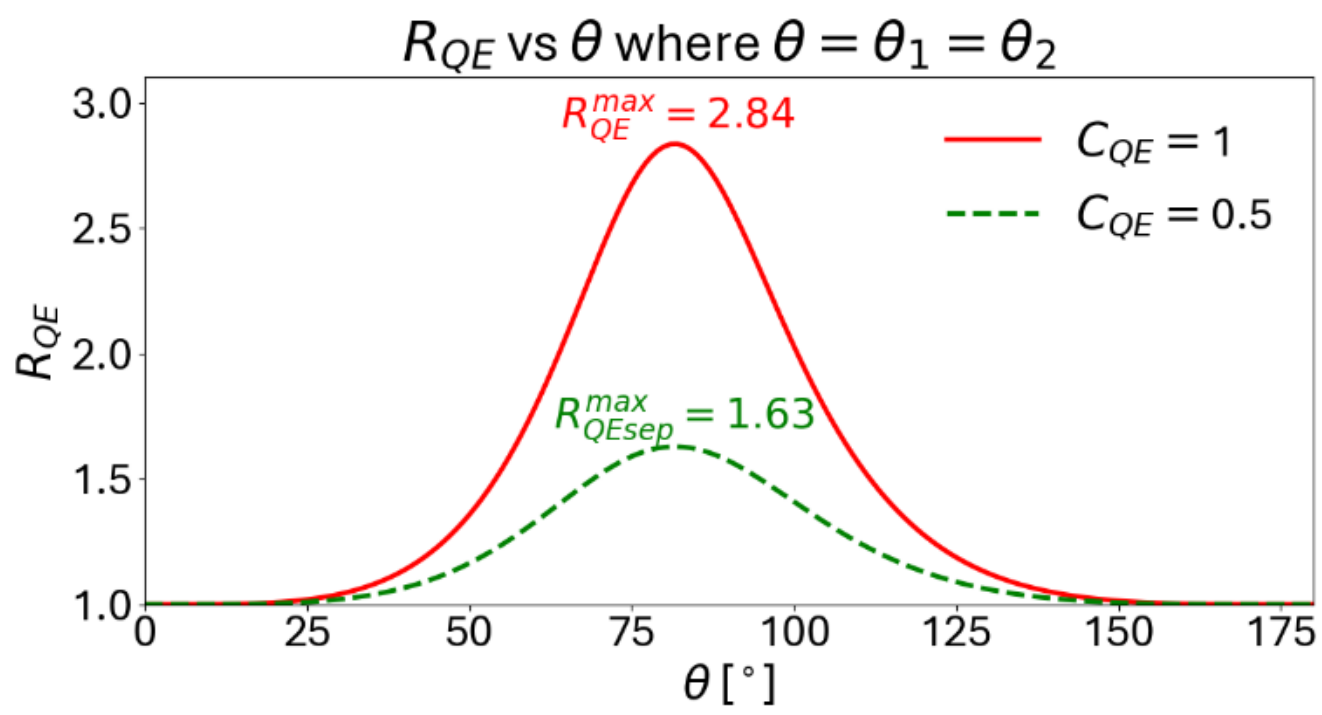

Fig. 5. Dependence of $R_{QE}(C_{QE} = 1)$ and $R_{QE}(C_{QE} = ½)$ as a function of $θ = θ_1 = θ_2$, calculated using equations 8, 9 and11.

## Tissue heterogeneity – the main challenge for detecting hypoxia with positronium

The mean ortho-positronium lifetime $\tau_{oPs}$, and hence also the ratio of 3γ-to-2γ decay rate ($R_{3\gamma/2\gamma}$ and $R_{oPs\text{-}3\gamma/2\gamma}$ ) are varying stronger due to the tissue type than due to the oxygen concentration changes [20]. Fig. 2C illustrates that the $\tau_{oPs}$ may vary even more than one nanosecond, while as it is shown in Fig. 6 (and later in detail in Fig. 8, Fig. 10, and Table II), the changes of $\tau_{oPs}$ (Δt) due to the different oxygen concentration between physoxic and hypoxic conditions are at the order of tens of picoseconds. Therefore, the absolute values of the ratio $R_{3\gamma/2\gamma}$ alone (as suggested in references [42, 44, 102, 103]), or the absolute value of $\tau_{oPs}$ alone (as postulated in references [20, 36, 37]) can be challenging to apply as a biomarker of hypoxia. At the first step perhaps more practical will be application of relative changes [44] or as we discuss in this manuscript, the simultaneous determination of $\tau_{oPs}$ and $R_{oPs\text{-}3\gamma/2\gamma}$ or determination of the degree of quantum entanglement $C_{QE}$.

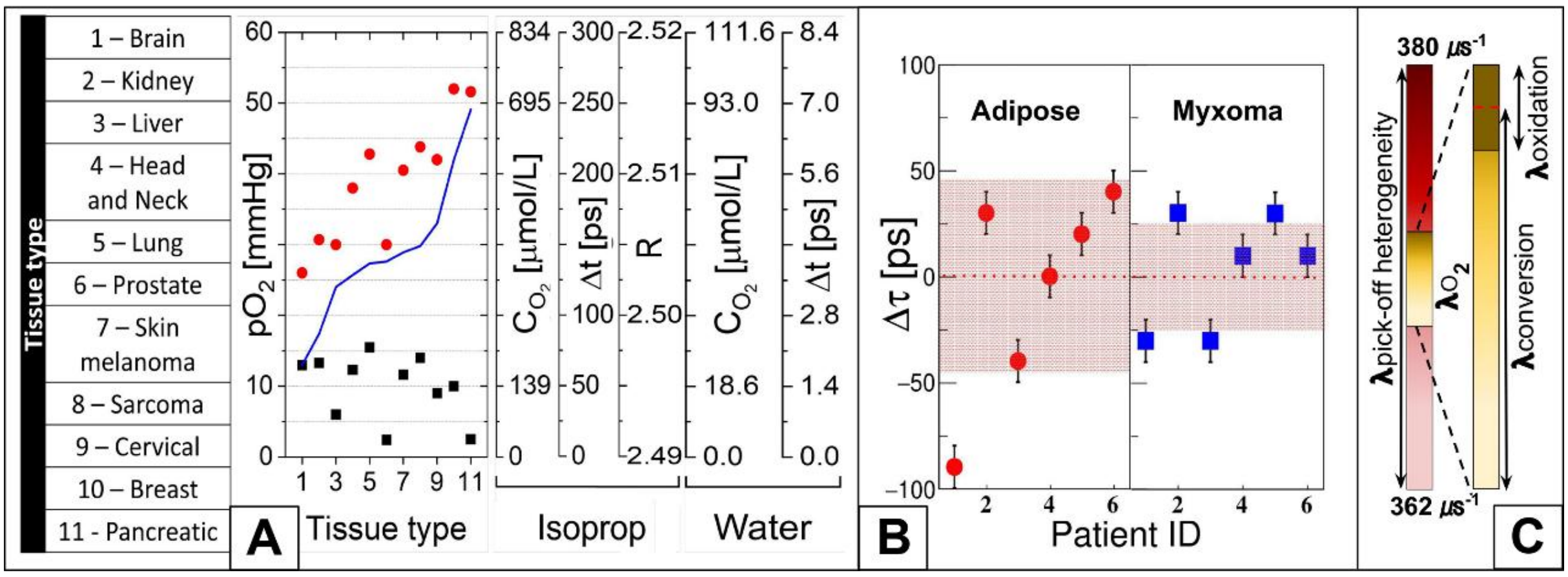

**Fig. 6. (A)** The partial pressure of oxygen molecules ($pO_2$) in healthy (red circles) and cancer (black squares) tissues [20]. Numbers indicating tissue type are explained in the legend shown in the left panel. The blue curve presents differences in $pO_2$ between healthy and neoplastic tissues. The right axes present concentration of oxygen ($C_{O2}$) in isopropanol and in water, and also changes of oPs mean lifetime $\tau_{oPs}$ with respect to $\tau_{oPs}$ at $pO_2$=0 ($\Delta t = \tau_{oPs}(pO_2) - \tau_{oPs}(0)$), as defined in equation (19) and shown in detail in Fig 8. Also expected influence of $pO_2$ on $R_{QE}$ for isopropanol is shown as an example of organic liquid (more details are in Fig. 12). The values are calculated based on [36, 37], assuming partial pressure of oxygen ($pO_2$) indicated on the left axis. **(B)** Differences of $\tau_{oPs}$ ($\Delta\tau = \tau_{average} - \tau_{ID}$) for adipose tissues (red circles) and cardiac myxoma tissues (blue squares) operated from six patients (ID) [46]. **(C)** Variation of $\lambda_{pick\text{-}off}$ due to heterogeneity of adipose tissue in patients compared to $\lambda_{O2} \approx 3.7\mu s^{-1}$ expected in case of 10 mmHg $O_2$ pressure in isopropanol.

Fig. 6A shows that (depending on the type of tissue) the partial pressure of oxygen in cancer tissues is smaller by about 10 mmHg to 50 mmHg with respect to healthy tissues. Decrease by 10 mmHg of oxygen partial pressure causes increase of $\tau_{oPs}$ only at the level of about one picosecond in water and ~50 ps in isopropanol. Such small effect is overwhelmed by $\tau_{oPs}$ variation due to tissue heterogeneity in different patients that are at the level of 100 ps, as shown in Fig. 6B for the adipose and cardiac myxoma tissues [46]. Therefore, it still must be established in medical practice whether the measurement of $\tau_{oPs}$ alone is sufficient for oxygenation estimation. This is because the variations in $\tau_{oPs}$ resulting from oxygen interaction with oPs in normal versus hypoxic conditions are smaller than those caused by tissue heterogeneity (Figs. 6B,C).

Measurement of $\tau_{oPs}$ tells us about the total oPs annihilation rate:

$$\lambda_{oPs} = \lambda_{oPs\text{-self}} + \lambda_{pick\text{-off}} + \lambda_{O_2} = \frac{1}{\tau_{oPs}} \tag{12}$$

where $\lambda_{oPs\text{-self}}$ describes the oPs self-annihilation. Knowing the values of $\lambda_{oPs\text{-self}}$, $\lambda_{pick\text{-off}}$ and dependence of $\lambda_{O2}$ on oxygen partial pressure ($pO_2$) for some organic liquids, we estimated (for details see section "Method 1: Dependence of $\tau_{oPs}$ and $R_{oPs\text{-}3\gamma/2\gamma}$ on $pO_2$") that the annihilation rate due to the oPs-$O_2$ interaction ($\lambda_{O2}$) changes between physoxic and hypoxic conditions only by a few $\mu s^{-1}$. Yet, variations of $\lambda_{pick\text{-off}}$ due to the tissue heterogeneity are at the level of 20 $\mu s^{-1}$ (Fig. 6C). Therefore, for assessing absolute values of $O_2$ concentration, it is necessary to distinguish contributions to 2γ annihilation from oPs-$O_2$ conversion and from oPs pick-off processes.

## Rate of conversion process as a function of the partial oxygen pressure $pO_2$

In order to estimate the rate of conversion processes as a function of partial pressure of oxygen we will first describe the dependence of the partial pressure of oxygen in liquid with respect to partial pressure in gas, next we will consider relation between partial pressure and concentration of oxygen in liquid, and we will also consider relation between concentration of oxygen and conversion rate in liquid. Further on we will consider dependence of $R_{oPs\text{-}3\gamma/2\gamma}$ and $\tau_{oPs}$ as a function of oxygen pressure in water, isopropanol, cyclohexane, isooctane, and adipose tissues. Finally, we will consider dependence on quantum entanglement witnesses $R_{QE}$ and $C_{QE}$ as a function of oxygen pressure in water, isopropanol, cyclohexane, isooctane, and adipose. We will derive formula for $pO_2$ as a function of $R_{oPs\text{-}3\gamma/2\gamma}$ and $\tau_{oPs}$ and propose phenomenological formula for $pO_2$ as a function of $R_{oPs\text{-}3\gamma/2\gamma}$.

In this manuscript following the Stepanov et al. [37, 104] and Shibuya at al. [36], taking into account that 1 mole of an ideal gas occupies 24.1 liters (L) at 20◦C, we will assume that the concentration of oxygen gas at 1atm ($pO_2$(1 atm) = 760 mmHg) is equal to:

$$C_{O2}(1\text{ atm}) = 1\text{ mol} / 24.1\text{ L} = 0.0415\text{ mol L}^{-1}. \tag{13}$$

The concentration of $O_2$ in liquid depends on the solubility that may be expressed in terms of the Ostwald coefficient (Ost_c) expressing the ratio of a given gas concentration in liquid to its concentration in the gas phase:

$$C_{O_2\text{-liquid}} = \text{Ost}_c \cdot C_{O_2\text{-gas}}. \tag{14}$$

Concentration of oxygen in liquid ($C_{O2\text{-liquid}}$ [μmol/L]) to the partial oxygen pressure ($pO_2$ [mmHg]) may be related as:

$$C_{O_2\text{-liquid}} = \text{Ost}_c \cdot \text{CP} \cdot pO_2, \tag{15}$$

where a proportionality coefficient (CP) relating concentration of oxygen in gas phase with normal atmosphere pressure at 1atm (760 mmHg) is equal to:

$$\text{CP} = C_{O2}(1\text{ atm}) / pO_2(1\text{ atm}) = 54.61\ \mu\text{mol L}^{-1}\text{ mmHg}^{-1}.$$

For $pO_2$ (partial pressure of oxygen), we use units mmHg because it is the standard unit in medical research and clinical practice for measuring gas pressures in the body, particularly in the context of hypoxia [3-7]. For molar concentration, we use μmol/L (micromoles per liter) as it provides a reasonable fit to the range of values encountered in medicine. Table I lists the $Ost_c$ coefficients for the water and organic liquids examined in this study, along with the corresponding $Ost_c$ * CP values provided for convenience.

Molecular oxygen $O_2$ dissolved in the tissue will contribute to the positronium annihilation rate mainly *via* processes of conversion ($oPs + O_2 \leftrightarrow pPs + O_2$) and oxidation ($Ps + O_2 \rightarrow e^+ + O_2^-$) [36, 37]. The annihilation rate due to conversion ($\lambda_{conv}$), and other processes (mainly oxidation) involving oxygen ($\lambda_{other}$) increases linearly with the concentration of dissolved oxygen molecules ($C_{O2}$) [36, 37]:

$$\lambda_{O2}(C_{O2}) = \lambda_{conv} + \lambda_{other} = k_{O2} \cdot C_{O2}\,, \quad (16)$$

and hence it also increases linearly with the partial pressure of oxygen ($pO_2$):

$$\lambda_{O2}(pO_2) = \lambda_{conv} + \lambda_{other} = k_{O2} \cdot Ost_C \cdot CP \cdot pO_2\,, \quad (17)$$

where $k_{O2}$ is a coefficient established experimentally [36, 37]. The values of $k_{O2}$ for water, adipose and organic liquids as isopropanol, cyclohexane and isooctane are given in Table I. The dependence of $\lambda_{O2}$ as a function of oxygen pressure ($pO_2$) and oxygen concentration ($C_{O2}$) for these substances is shown in Fig. 7. The result indicates that the positronium quenching effect (shortening of lifetime thus increasing decay rate) by molecular oxygen is most pronounced in isooctane, followed by a decreasing trend through cyclohexane, isopropanol, and adipose tissue, with the least effect observed in water.

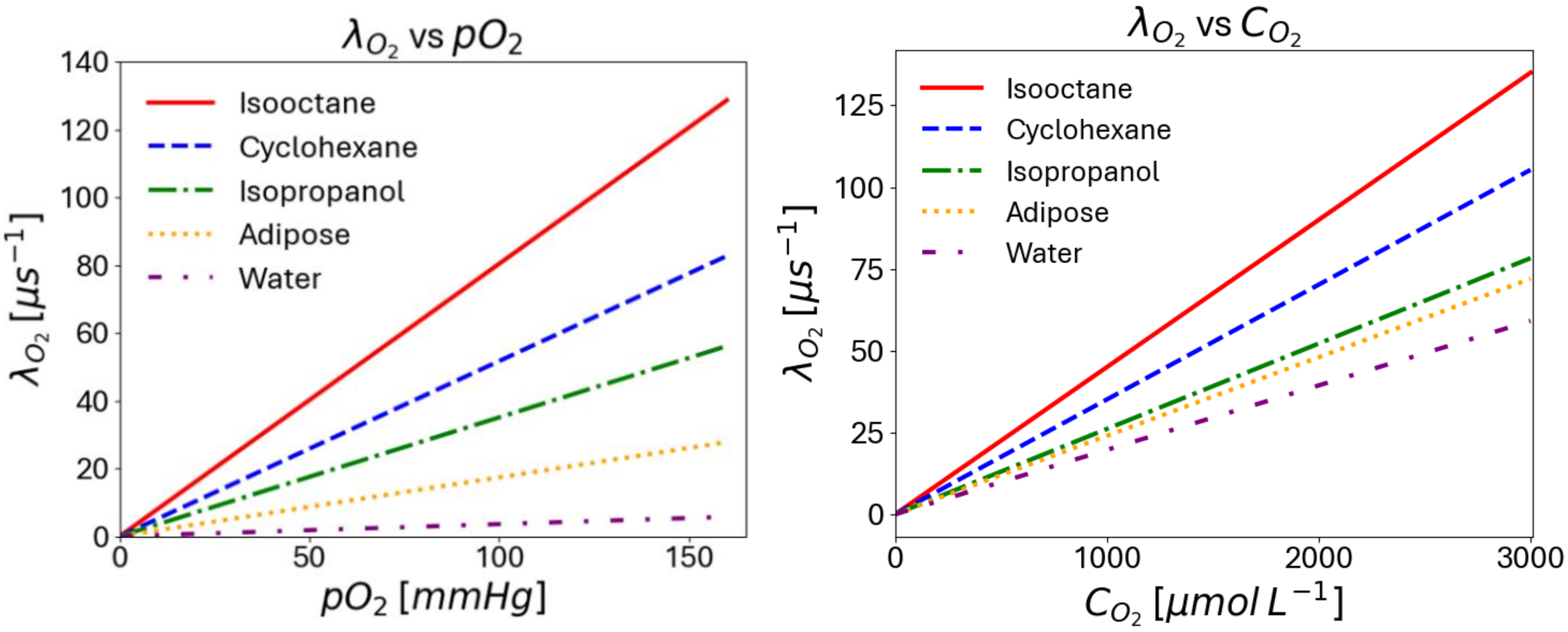

Fig. 7. (left) oPs decay rate due to the interactions with oxygen ($\lambda_{O2}$) as a function of oxygen pressure $pO_2$. The range of $pO_2$ from 0 to 60 mmHg is relevant for the tissue (see Fig. 6A). (right) $\lambda_{O2}$ as a function of oxygen concentration ($C_{O2}$). Calculations were performed using formulas 16 and 17. The values of the parameters $k_{O2}$, $Ost_c$ and CP used in the calculations are given in Table I. Concentration of oxygen in water at standard atmospheric pressure of 760 mmHg and 20 °C would be equal to 1386 μmol $L^{-1}$.

**Table I**. Values of coefficients $Ost_C$, $k_{O2}$, $\lambda_{pick\text{-}off}$, $f_d$, $f_{pPs}$, and $f_{oPs}$, used in this manuscript to estimate the values of $R_{oPs\text{-}3\gamma/2\gamma}$ and $\tau_{oPs}$ as well as to estimate the degree of quantum entanglement $R_{QE}$ and $C_{QE}$ as a function of partial oxygen pressure in water, isopropanol, cyclohexane, isooctane and in the adipose tissue. The values in this table are from references [13, 20, 36, 37, 104-111] as it is explained in detail in the text. In this manuscript, we do not discuss experimental uncertainties of used parameters. Here, we focus on the method itself and aim to establish the order of magnitude of changes in $\lambda_{O2}$, $\tau_{oPs}$, $R_{oPs\text{-}3\gamma/2\gamma}$, $C_{QE,}$ and $R_{QE}$ between physoxic and hypoxic conditions.

| **Substance** | **$Ost_C$** | **$Ost_C$ * CP**<br>**μmol $L^{-1}$ $mmHg^{-1}$** | **$k_{O2}$**<br>**$\mu s^{-1}\mu mol^{-1}L$** | **$\lambda_{pick\text{-}off}$**<br>**$\mu s^{-1}$** | **$f_d$** | $f_{Ps}$ **$f_{pPs}$** | $f_{Ps}$ **$f_{oPs}$** |
|---|---|---|---|---|---|---|---|
| Water | 0.0334 | 1.824 | 0.0197 | 513 | 0.44 | 0.31 | 0.25 |
| Isopropanol | 0.2463 | 13.45 | 0.0261 | 260 | 0.558 | 0.217 | 0.225 |
| Cyclohexane | 0.27 | 14.74 | 0.0351 | 310 | 0.389 | 0.25 | 0.361 |
| Isooctane | 0.362 | 19.77 | 0.0450 | 250 | 0.323 | 0.246 | 0.431 |
| Adipose | 0.133 | 7.241 | 0.024 | 368 | 0.613 | 0.172 | 0.215 |

The values of $Ost_C$ parameter shown in Table I are from references [37, 104-108]. The values $Ost_C$ * CP represent the ratio of a chemical's aqueous-phase concentration to its equilibrium partial pressure in the gas phase and are known as Henry's law solubility constants [109, 110]. In this table, $Ost_C$ value for adipose is given for 37 °C [108], while for other substances $Ost_C$ values at 20 °C are given. $Ost_C$ coefficient is decreasing with the increase of temperature. Typically, it decreases by one to a few per cent per 1 °C increase in temperature [108], e.g. for water $Ost_C$ = 0.027 at 37 °C [108]. The coefficient $k_{O2}$ = 0.0197± 0.0006 $\mu mol^{-1}\mu s^{-1}L$ for water is calculated [20] as weighted mean of $k_{O2}$ = 0.0204 ± 0.0008 $\mu mol^{-1}\mu s^{-1}L$ [36] and $k_{O2}$ = 0.0186 ± 0.0010 $\mu mol^{-1}\mu s^{-1}L$ [37]. Coefficient $k_{O2}$ ~ 0.024 $\mu s^{-1}\mu mol^{-1}L$ for adipose (oil) is estimated based on Fig. 3 in reference [37] showing dependence of $k_{O2}$ on the reversed viscosity η. It is obtained assuming 1/η ≈ 0 (because of large value of η for olive oil, η = 56.2 cP [111]). $\lambda_{pick\text{-}off}$ = 513 +- 1.6 $\mu s^{-1}$ for water is calculated as a weighted mean of 512.8 +- 1.6 $\mu s^{-1}$ [36] and 550 +-20 $\mu s^{-1}$ [37]. The values of $f_{d,}$. $f_{pPs}$. $f_{oPs}$ for adipose are taken from reference [13], and for water, isopropanol, cyclohexane and isooctane are taken from reference [37].

## Ortho-positronium lifetime as a function of the partial oxygen pressure $pO_2$

Knowing the values of $\lambda_{oPs\text{-}self}$, $\lambda_{pick\text{-}off}$, and knowing dependence of $\lambda_{O2}$ on $O_2$ concentration we can now express the rate ($\lambda_{oPs}$) and hence the mean lifetime of ortho-positronium ($\tau_{oPs}$) in a given substance by:

$$\frac{1}{\tau_{\text{oPs}}}(C_{O2}) = \lambda_{\text{oPs}}(C_{O2}) = \lambda_{\text{oPs-self}} + \lambda_{\text{pick-off}} + \lambda_{O_2}(C_{O2}) = \lambda_{\text{oPs-self}} + \lambda_{\text{pick-off}} + k_{O_2} \cdot \mathrm{C}_{O2} \tag{18}$$

Combining equations 16, 17, and 18 we may define the change of the mean oPs lifetime (Δt) in a given substance as a function of oxygen concentration ($C_{O2}$) and as a function of oxygen partial pressure ($pO_2$) as follows:

$$\Delta t(pO_2) = \tau_{oPs}(0) - \tau_{oPs}(pO_2) = \frac{1}{\lambda_{oPs\text{-self}} + \lambda_{\text{pick-off}}} - \frac{1}{\lambda_{\text{oPs-self}} + \lambda_{\text{pick-off}} + k_{O_2} \cdot \mathrm{Ost_C} \cdot CP \cdot p_{O_2}} \tag{19}$$

$$\Delta t(C_{O2}) = \tau_{oPs}(0) - \tau_{oPs}(C_{O2}) = \frac{1}{\lambda_{oPs\text{-self}} + \lambda_{\text{pick-off}}} - \frac{1}{\lambda_{oPs\text{-self}} + \lambda_{\text{pick-off}} + k_{O_2} \cdot C_{O2}} \tag{20}$$

Fig. 8 shows the dependence of Δt on $C_{O2}$ and $pO_2$. The observed tendency is the same as for the decay rate in Fig. 7. The changes of $\tau_{oPs}$ due to the molecular oxygen are most pronounced in isooctane, followed by a decreasing trend through cyclohexane, isopropanol, and adipose tissue, with the least effect observed in water.

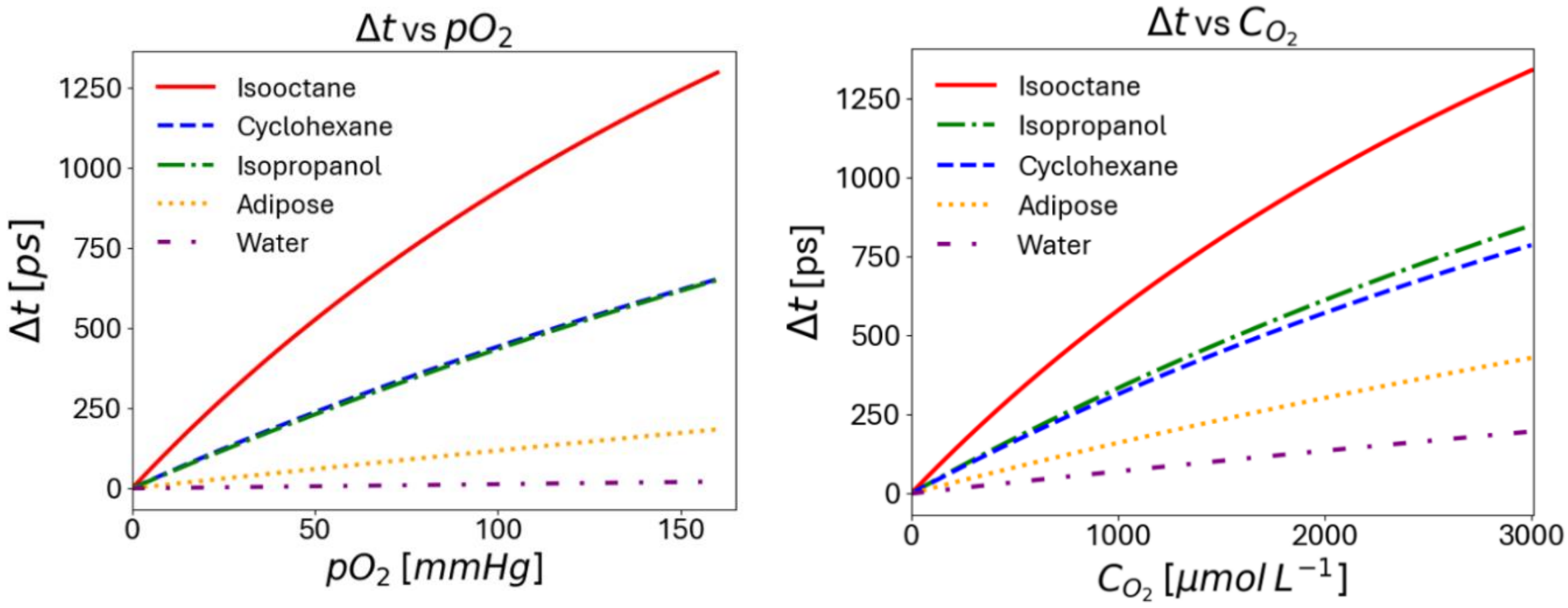

Fig. 8. (left) Changes of the mean oPs lifetime (Δt) as a function of oxygen pressure $pO_2$ with respect to mean oPs lifetime at $pO_2$= 0. The range of $pO_2$ from 0 to 60 mmHg is relevant for the tissue (see Fig. 6A). (right) Δt as a function of oxygen concentration ($C_{O2}$). Calculations were performed using formulas 19 and 20. The values of the parameters $k_{O2}$, $Ost_c$, CP, and $\lambda_{pick\text{-}off}$ used in the calculations are given in Table I. The value of $\lambda_{o\text{-}Ps\text{-}self}$ is 7.0401 $\mu s^{-1}$ [64]. Concentration of oxygen in water at atmospheric pressure of 760 mmHg and 20 °C would be equal to 1386 μmol $L^{-1}$.

## The ratio of 2γ pick-off annihilations to all 2γ annihilations as a function of the partial oxygen pressure.

In this section we calculate how the 2γ fraction due to pick-off annihilations depends on the oxygen concentration.

In this work, to estimate the degree of QE, we assume that the photons (2γ) from pick-off annihilations are not QE, whereas photons (2γ) from other processes are maximally QE. This serves as an initial hypothesis for estimating the expected effect. Therefore, of particular importance is the ratio of 2γ pick-off annihilations to all 2γ annihilations, expressed as:

$$\alpha = \frac{f_{2\gamma\text{-pick-off}}}{f_{2\gamma\text{-all}}}, \tag{21}$$

where

$$f_{2\gamma\text{-all}} = f_{2\gamma\text{-pick-off}} + f_{2\gamma\text{-direct}} + f_{2\gamma\text{-pPs-self}} + f_{2\gamma\text{-}O_2}, \tag{22}$$

with $f_{2\gamma\text{-all}}$, $f_{2\gamma\text{-pick-off}}$, $f_{2\gamma\text{-direct}}$, $f_{2\gamma\text{-pPs-self}}$, $f_{2\gamma\text{-O2}}$ denoting total fraction of 2γ annihilations, fraction of annihilations via 2γ pick-off process, fraction of annihilations via direct 2γ annihilation, fraction of annihilations via para-positronium 2γ annihilations, and fraction of oPs annihilations into 2γ due to interaction with oxygen molecule. These fractions may be expressed as:

$$f_{2\gamma\text{-pick-off}} = f_{\mathrm{Ps}} f_{\mathrm{oPs}} \alpha_2 \frac{\lambda_{\text{pick-off}}}{\lambda_{\mathrm{oPs}}} + f_{\mathrm{Ps}} f_{\mathrm{pPs}} \alpha_2 \frac{\lambda_{\text{pick-off}}}{\lambda_{\mathrm{pPs}}} \approx f_{\mathrm{Ps}} f_{\mathrm{oPs}} \alpha_2 \frac{\lambda_{\text{pick-off}}}{\lambda_{\mathrm{oPs}}} \tag{23}$$

$$f_{2\gamma\text{-direct}} = f_d \beta_2\,, \tag{24}$$

$$f_{2\gamma\text{-pPs-self}} = f_{\mathrm{Ps}} f_{\mathrm{pPs}} \frac{\lambda_{\text{pPs-self}}}{\lambda_{\mathrm{pPs}}}\,, \tag{25}$$

$$f_{2\gamma\text{-O}_2} = f_{\mathrm{Ps}} f_{\mathrm{oPs}} \frac{\beta_2 \lambda_{O_2\text{-other}} + \lambda_{O_2\text{-conv}}}{\lambda_{\mathrm{oPs}}} + f_{\mathrm{Ps}} f_{\mathrm{pPs}} \frac{\beta_2 \lambda_{O_2\text{-other}}}{\lambda_{\mathrm{pPs}}} \tag{26}$$

$$f_{2\gamma\text{-O}_2} \approx f_{\mathrm{Ps}}\, f_{\mathrm{oPs}} \left( \beta_2 \frac{\omega}{1+\omega} + \frac{1}{1+\omega} \right) \frac{\lambda_{O_2}}{\lambda_{\mathrm{oPs}}} + f_{\mathrm{Ps}} f_{\mathrm{pPs}}\, \beta_2 \frac{\omega}{1+\omega} \frac{\lambda_{O_2}}{\lambda_{\mathrm{pPs}}} \tag{27}$$

$$f_{2\gamma\text{-O}_2} \approx f_{\mathrm{Ps}}\, f_{\mathrm{oPs}} \frac{\lambda_{O_2}}{\lambda_{\mathrm{oPs}}} \tag{28}$$

where $f_d$, $f_{Ps}$, $f_{pPs}$, $f_{oPs}$, $\alpha_2$, $\beta_2$ are defined in the caption and description of Fig. 3, ω is defined in equation (4), values of $\lambda_{\text{pick-off}}$ are listed in Table I, the dependence of $\lambda_{O2}$ on the partial oxygen pressure $pO_2$ is given in equation (17), and the $\lambda_{oPs}$ and $\lambda_{pPs}$ are defined in equations (1), (2) and (12). Using the values of these parameters and applying the above equations we can determine α as a function of $pO_2$. The result is shown in Fig. 9. Among the studied substances the largest variation of the value of parameter α as a function of oxygen pressure $pO_2$ is observed for isooctane. Following this, in order of decreasing changes, are cyclohexane, isopropanol, adipose tissue, and water. Although the ratio of 2γ pick-off annihilations to all 2γ annihilations in water is relatively high, its changes as a function of oxygen pressure are small due to the small solubility of oxygen compared to other considered substances (compare Ostwald coefficients in Table I).

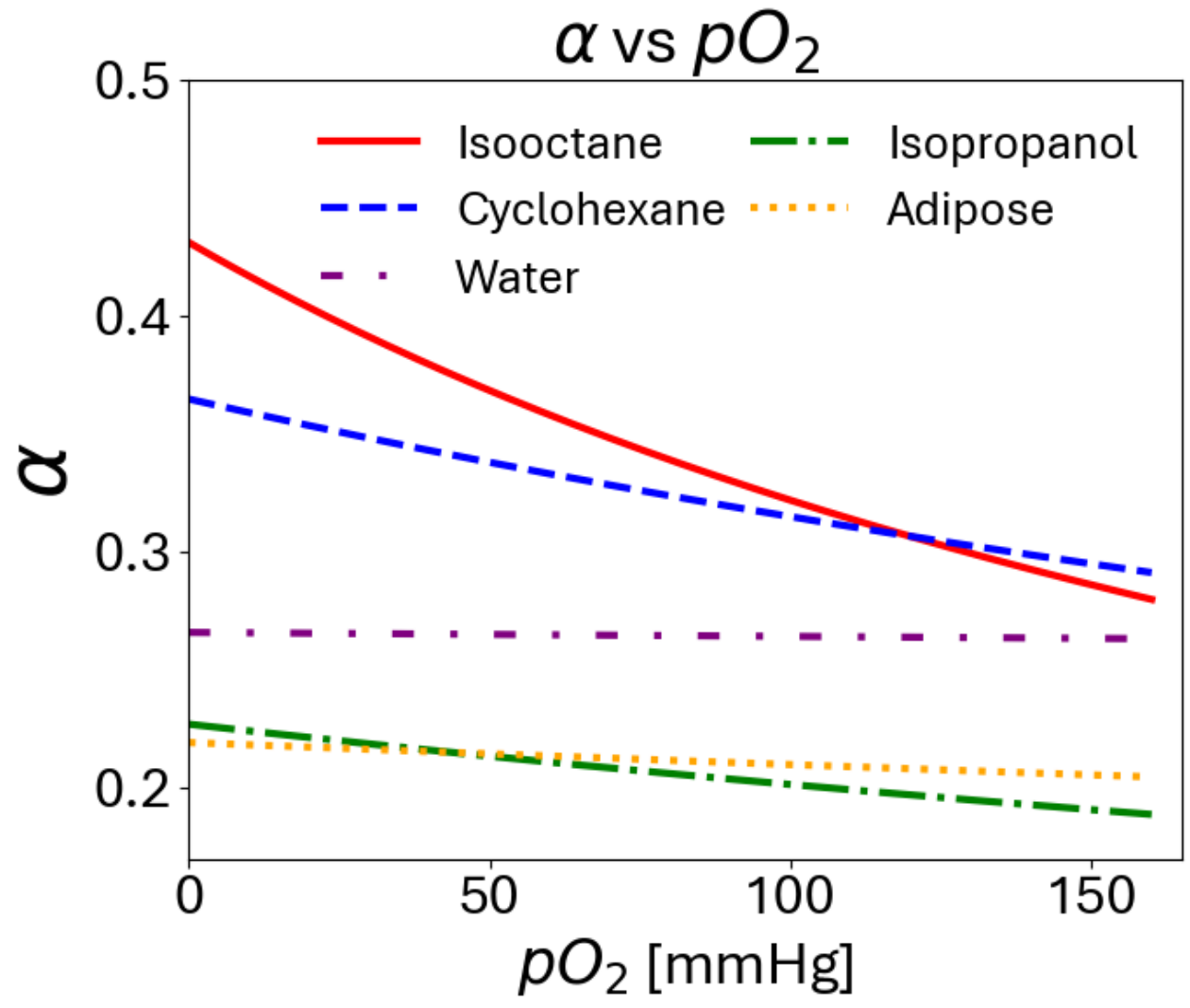

Fig. 9. Parameter α as a function of oxygen pressure $pO_2$ determined for water, isopropanol, cyclohexane, isooctane, and adipose. The range of $pO_2$ from 0 to 60 mmHg is relevant for the tissue (see Fig. 6A). Calculations were performed using formulas 21,22,23,24,25,26, as well as 1,2,3,4,5,12 and 17. The values of the parameters $k_{O2}$, $Ost_c$, CP, and $\lambda_{\text{pick-off}}$ used in the calculations are given in Table I. The value of $\lambda_{\text{o-Ps-self}}$ is 7.0401 $\mu s^{-1}$ [64] and the value of $\lambda_{\text{p-Ps-self}}$ is 7990.9 $\mu s^{-1}$ [63]. The value of ω = 0.15 is assumed.

## Method 1: Dependence of $\tau_{oPs}$ and $R_{oPs\text{-}3\gamma/2\gamma}$ on $pO_2$

In 2004 Kacperski and Spyrou [42, 102, 103] suggested the ratio of 3γ-to-2γ decay rates as a possible indicator of hypoxia and proposed a method to reconstruct the position of $e^+e^-$ annihilation into three photons based on the measurement of photons' energies and positions of interactions with high energy-resolution semiconductor detectors (such as e.g. CdZnTe). Yet, it was concluded that due to the low fraction of 3γ events this method cannot provide the sensitivity needed to detect changes in the tissue of the dissolved oxygen concentrations [45]. Moreover, the application of semiconductor detectors would make PET systems even more expensive than those based on crystal scintillators.

The situation changed with the elaboration and first demonstration of 3γ imaging by means of plastic scintillators [23] making the application of ratio of 3γ-to-2γ decay rates more realistic in hospitals and by recent developments of 3γ imaging by means of crystal detectors [112-116].

Here we discuss the possibilities of determining $pO_2$ by simultaneous measurement of the ratio of 3γ-to-2γ decay rates ($R_{oPs\text{-}3\gamma/2\gamma}$), and oPs lifetime ($\tau_{oPs}$).

Taking into account equations (5), (7), (12) and (17), we can express $R_{oPs\text{-}3\gamma/2\gamma}$ as:

$$R_{oPs-3\gamma/2\gamma} = \frac{\lambda_{oPs-self} + \frac{\alpha_3}{\tau_{oPs}} - \alpha_3\lambda_{oPs-self} + \left(\beta_3 \frac{\omega}{1+\omega} - \alpha_3\right) k_{O_2} \cdot Ost_c \cdot CP \cdot pO_2}{\frac{\alpha_2}{\tau_{oPs}} - \alpha_2\lambda_{oPs-self} + \left(\beta_2 \frac{\omega}{1+\omega} + \frac{1}{1+\omega} - \alpha_2\right) k_{O_2} \cdot Ost_c \cdot CP \cdot pO_2} \quad (29)$$

Based on equations (12) and (17) we can relate $\tau_{oPs}$ and $pO_2$ with the following formula:

$$\tau_{oPs}(p_{O_2}) = \frac{1}{\lambda_{oPs-self} + \lambda_{pick-off} + k_{O_2} \cdot Ost_c \cdot CP \cdot pO_2} \quad (30)$$

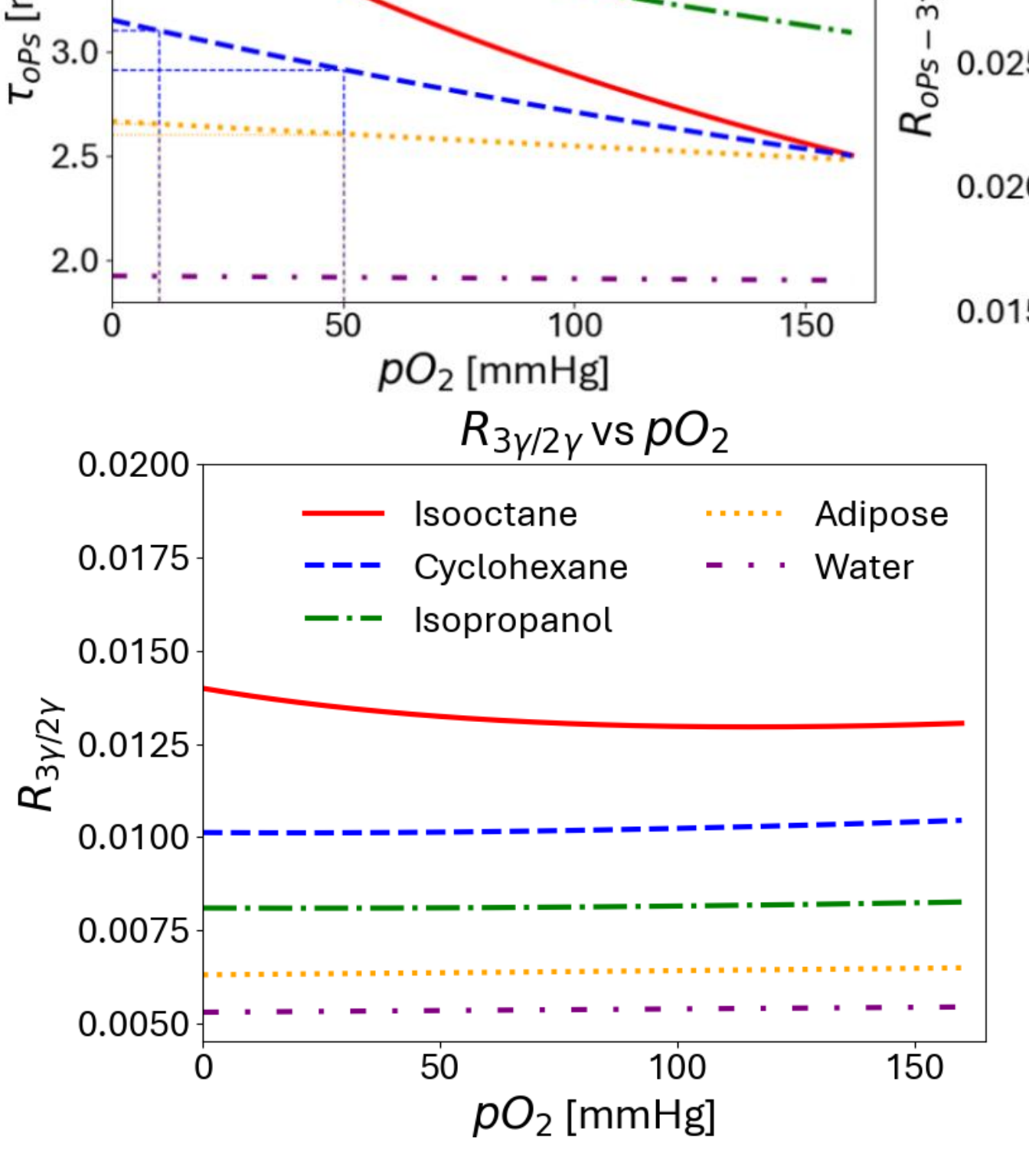

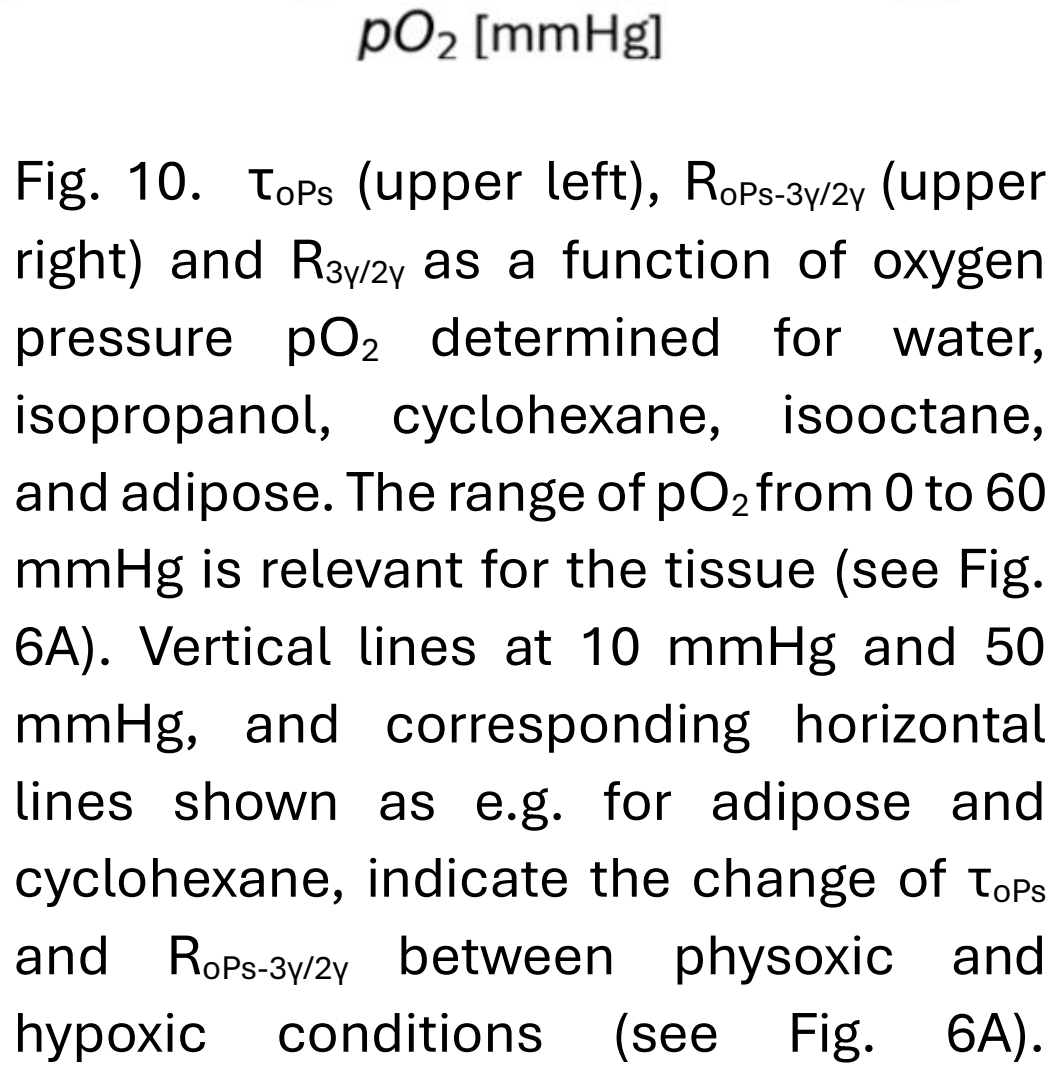

Fig. 10. $\tau_{oPs}$ (upper left), $R_{oPs\text{-}3\gamma/2\gamma}$ (upper right) and $R_{3\gamma/2\gamma}$ as a function of oxygen pressure $pO_2$ determined for water, isopropanol, cyclohexane, isooctane, and adipose. The range of $pO_2$ from 0 to 60 mmHg is relevant for the tissue (see Fig. 6A). Vertical lines at 10 mmHg and 50 mmHg, and corresponding horizontal lines shown as e.g. for adipose and cyclohexane, indicate the change of $\tau_{oPs}$ and $R_{oPs\text{-}3\gamma/2\gamma}$ between physoxic and hypoxic conditions (see Fig. 6A).

Calculations were performed using formulas 29, 30 and 6. The dependence of $\lambda_{O2}$ on the partial oxygen pressure $pO_2$ is given in equation (17), and the $\lambda_{oPs}$ and $\lambda_{pPs}$ are defined in equations 1, 2 and 12. The values of the parameters $k_{O2}$, $Ost_c$, CP, and $\lambda_{pick\text{-}off}$, $f_d$, $f_{Ps}$, $f_{pPs}$, $f_{oPs}$, used in the calculations are given in Table I. The value of $\lambda_{o\text{-}Ps\text{-}self}$ is 7.0401 $\mu s^{-1}$ [64]. In calculations we assume (see text below Fig. 3) that $\alpha_2 = 378/379$, $\alpha_3 = 1/379$, $\beta_2 = 371.3/372.3$, $\beta_3 = 1/372.3$. The value of $\omega = 0.15$ is used.

Fig. 10 shows the calculated dependence of $\tau_{oPs}$, $R_{oPs\text{-}3\gamma/2\gamma}$ and $R_{3\gamma/2\gamma}$ as a function of $pO_2$ for water, isopropanol, cyclohexane, isooctane, and adipose. The values of $R_{oPs\text{-}3\gamma/2\gamma}$ and $R_{3\gamma/2\gamma}$ at $pO_2 = 0$, as well as the difference in $\tau_{oPs}$ and $R_{oPs\text{-}3\gamma/2\gamma}$ calculated for physoxic and hypoxic conditions is shown in Table II. The calculations were performed assuming $pO_2 = 10$ mmHg for hypoxia and 50 mmHg for physoxia. In order to distinguish between physoxic and hypoxic conditions the precision of determining $\tau_{oPs}$ must be few times better than values shown in Table II. For adipose, a precision would need to be good enough to see difference of $\tau_{oPs}$ smaller than 48 ps, and difference in $R_{oPs\text{-}3\gamma/2\gamma}$ smaller than 0.0004. Fig. 10 shows also that, as noted earlier in this manuscript, the changes in $R_{oPs\text{-}3\gamma/2\gamma}$ as a function of oxygen pressure are much larger than changes in $R_{3\gamma/2\gamma}$.

**Table II**. Differences between the values of $\lambda_{O2}$, $\tau_{oPs}$, $R_{oPs\text{-}3\gamma/2\gamma}$, $C_{QE}$ and $R_{QE}$ between physoxic and hypoxic conditions. The values as a function of $pO_2$ are shown in Fig. 7, Fig. 10, and Fig. 12. In this manuscript, we do not discuss uncertainties arising from experimental errors in the parameters used. Instead, the aim of this study is to establish the required order of magnitude of precision for investigating changes in $\lambda_{O2}$, $\tau_{oPs}$, $R_{oPs\text{-}3\gamma/2\gamma}$, $C_{QE}$, and $R_{QE}$ between physoxic and hypoxic conditions. The values for $R_{QE}$ are calculated for $\theta = \theta_1 = \theta_2 = 81.7°$.
$\Delta\lambda_{O2}$, $\Delta\tau_{oPs}$, $\Delta R_{oPs\text{-}3\gamma/2\gamma}$, $\Delta C_{QE}$, $\Delta R_{QE}$ are defined as follows:
$\Delta R_{oPs\text{-}3\gamma/2\gamma} = R_{oPs\text{-}3\gamma/2\gamma}(pO_2 = 10\text{ mmHg}) - R_{oPs\text{-}3\gamma/2\gamma}(pO_2 = 50\text{ mmHg})$;
$\Delta\tau_{oPs} = \tau_{oPs}(pO_2 = 10\text{ mmHg}) - \tau_{oPs}(pO_2 = 50\text{ mmHg})$;
$\Delta C_{QE} = C_{QE}(pO_2 = 50\text{ mmHg}) - C_{QE}(pO_2 = 10\text{ mmHg})$;
$\Delta R_{QE} = R_{QE}(pO_2 = 50\text{ mmHg}) - R_{QE}(pO_2 = 10\text{ mmHg})$;
$\Delta\lambda_{O2} = \lambda_{O2}(pO_2 = 50\text{ mmHg}) - \lambda_{O2}(pO_2 = 10\text{ mmHg})$.
In the last column the calculated values of $C_{QE}$ for $pO_2 = 0$ are given.

| **Substance** | $\Delta\lambda_{O2}$ | $\Delta\tau_{oPs}$ | $\Delta R_{oPs-3\gamma/2\gamma}$ | $\Delta \boldsymbol{C_{QE}}$ | $\Delta \boldsymbol{R_{QE}}$ | $C_{QE}$ ($pO_2$=0) | $R_{3\gamma/2\gamma}$ ($pO_2$= 0) | $R_{oPs\text{-}3\gamma/2\gamma}$ ($pO_2$=0) |
|---|---|---|---|---|---|---|---|---|
| | Fig. 7 | Fig. 10 | Fig. 10 | Fig. 12 | Fig. 12 | Fig. 12 | Fig. 10 | Fig. 10 |
| Water | 1.4 $\mu s^{-1}$ | 5 ps | 0.000045 | 0.0003 | 0.0009 | 0.867 | 0.0053 | 0.0164 |
| Isopropanol | 14.0 $\mu s^{-1}$ | 182 ps | 0.0015 | 0.0054 | 0.0156 | 0.886 | 0.0081 | 0.0298 |
| Cyclohexane | 20.7 $\mu s^{-1}$ | 187 ps | 0.0015 | 0.0106 | 0.0276 | 0.818 | 0.0101 | 0.0254 |
| Isooctane | 35.6 $\mu s^{-1}$ | 444 ps | 0.0036 | 0.0243 | 0.0614 | 0.784 | 0.0140 | 0.0309 |
| Adipose | 7.0 $\mu s^{-1}$ | 48 ps | 0.0004 | 0.0019 | 0.0055 | 0.890 | 0.0063 | 0.0218 |

## Estimating $pO_2$ based on $\tau_{oPs}$ and $R_{oPs\text{-}3\gamma/2\gamma}$

In the previous sections we have considered how oxygen pressure influences oPs lifetime $\tau_{oPs}$ and decay rate ratio $R_{oPs-3\gamma/2\gamma}$, and we derived formulae describing dependence of $\tau_{oPs}$ and $R_{oPs-3\gamma/2\gamma}$ on molecular oxygen pressure $pO_2$. In clinical practice we will need to determine $pO_2$ based on the measured $\tau_{oPs}$ and $R_{oPs-3\gamma/2\gamma}$. Therefore, in this section we will introduce and discuss formulas enabling calculation of $pO_2$ based on the measurement of $\tau_{oPs}$ and $R_{oPs-3\gamma/2\gamma}$.

The formula for calculating $pO_2$ as a function of $\tau_{oPs}$ and $R_{oPs-3\gamma/2\gamma}$, obtained by solving equation (29) for $pO_2$, is given by:

$$pO_2 = \frac{(1 + R_{oPs-3\gamma/2\gamma})\alpha_2\lambda_{oPs-self} + \frac{\alpha_3 - \alpha_2 R_{oPs-3\gamma/2\gamma}}{\tau_{oPs}}}{\left(R_{oPs-3\gamma/2\gamma}\left(\beta_2\frac{\omega}{1+\omega} + \frac{1}{1+\omega} - \alpha_2\right) - \left(\beta_3\frac{\omega}{1+\omega} - \alpha_3\right)\right)k_{O_2}\cdot Ost_c \cdot CP} \quad (31)$$

In practice the above formula will be applicable provided that the parameters $Ost_c$, ω and $k_{O2}$ will be established for the tissue of interest.

A more practical approach would be to introduce a phenomenological formula reflecting the fact that as $pO_2$ grows, both the $R_{oPs-3\gamma/2\gamma}$ ratio and $\tau_{oPs}$ decrease:

$$pO_2 = p_1 + \frac{p_2}{R_{oPs-3\gamma/2\gamma}} + \frac{p_3}{\tau_{oPs}} + \frac{p_4}{R_{oPs-3\gamma/2\gamma}\cdot\tau_{oPs}} \quad (32)$$

where p1, p2, p3, and p4 are effective parameters that need to be determined experimentally for each tissue.

## Method 2: Dependence of degree of quantum entanglement on $pO_2$

In the case when oPs annihilates *via* interactions with an electron from the surrounding atoms (pick-off process in Fig. 2B), the initial state will be a mixture of electron-positron states with many possible quantum numbers. Hence, we may anticipate that the shape of the φ-distribution $F(\theta_1, \theta_2, C_{QE}, \varphi)$ will be a combination of solid and dashed curves (Fig. 4B) with the degree of entanglement $C_{QE}$ between 0.5 and 1 (corresponding to $R_{QE}$ between $R_{QE}$_sep and $R_{QE}$_max (Fig. 5)). In turn, for the conversion process (oPs + $O_2$ → pPs + $O_2$ → 2γ + $O_2$) in which photons originate from pPs → 2γ decay, we anticipate that the 2γ state is maximally entangled resulting in the strong correlation in the φ-distribution ($C_{QE}$ = 1 in equation 8) indicated by red curve in Fig. 4B. Thus, we expect differences of the φ-distribution depending on the degree of the entanglement of emitted photons, which in turn may allow one to determine the contributions originating from oPs pick-off and from oPs conversion processes.

In general, we may assume that photons are maximally quantum entangled for an $\alpha_m$ fraction of 2γ annihilations, and the remaining (1-$\alpha_m$) fraction of the 2γ annihilations produced photons propagating independently of each other. Without loss of generality the measured φ-distribution F(φ) may be decomposed as:

$$F(\theta_1, \theta_2, \varphi, \alpha_m) = \alpha_m F(\theta_1, \theta_2, \varphi, C_{QE} = 1) + (1 - \alpha_m)F(\theta_1, \theta_2, \varphi, C_{QE} = 1/2). \quad (33)$$

Employing (8), (9) and (33) we get:

$$C_{QE}(\alpha_m) = \frac{\alpha_m + 1}{2} \quad (34)$$

and employing (33) in (11) we obtain:

$$R_{QE}(\theta_1,\theta_2,\alpha_m) = \frac{1+\alpha_m\frac{R_{QEmax}-1}{R_{QEmax}+1}+(1-\alpha_m)\frac{R_{QEsep}-1}{R_{QEsep}+1}}{1-\alpha_m\frac{R_{QEmax}-1}{R_{QEmax}+1}-(1-\alpha_m)\frac{R_{QEsep}-1}{R_{QEsep}+1}}, \quad (35)$$

where we denoted:

$$R_{QEmax} = R_{QE}(\theta_1,\theta_2, C_{QE}=1) \text{ and } R_{QEsep} = R_{QE}(\theta_1,\theta_2, C_{QE}=1/2). \quad (36)$$

Here, parameter $\alpha_m$ was introduced as a fraction of 2γ annihilations with maximally entangled photons, and (1- $\alpha_m$) as fraction of separable photons. However, in the matter (in particular, in the tissue), we may expect a spectrum of different values for the degree of entanglement. Hence, $\alpha_m$ should be treated as a phenomenological parameter, which accounts, on average, for the observed $F(\theta_1,\theta_2,\varphi,\alpha_m)$ distribution.

It is worth noting that $R_{QE}$ and $C_{QE}$ are related to each other, but $R_{QE}$ depends on the Compton scattering angles, while $C_{QE}$ is a measure of the degree of quantum entanglement that is independent of the scattering angles.

For example, for $\theta = \theta_1 = \theta_2 = 81.7°$ with $R_{QEmax}$ = 2.84 and $R_{QEsep}$ = 1.63 equation (35) reads:

$$R_{QE}(\theta_1=\theta_2=81.7°,\alpha_m) \approx \frac{1.239+0.239\,\alpha_m}{0.760-0.239\,\alpha_m}$$

and using equation (34) we obtain:

$$R_{QE}(\theta_1=\theta_2=81.7°, C_{QE}) \approx \frac{1.001+0.478\,C_{QE}}{0.999-0.478\,C_{QE}}$$

Fig. 11. $R_{QE}$ and $C_{QE}$ as a function of $\alpha_m$ (left), and as a function of α (right), determined for few exemplary scattering angles $\theta = \theta_1 = \theta_2$. Calculations were performed using formulas 35, 36, 8, 9, 11 and 37. Lines for $C_{QE}$ are calculated using equations 34 (left) and 38 (right).

We hypothesize that the value of $\alpha_m$ changes in response to variations in $pO_2$. We will elaborate on this dependence in the next section. The left panel of Fig. 11 shows the dependence of $R_{QE}$ on $\alpha_m$ for a few chosen scattering angles, $\theta = \theta_1 = \theta_2$. As seen in Fig. 11, the dependence of $R_{QE}$ on $\alpha_m$ is most pronounced for $\theta = \theta_1 = \theta_2 = 81.7°$ and becomes very weak for angles larger than 130° and smaller than 30°, which is also evident in Fig. 5. The θ-independent $C_{QE}$ as a function of $\alpha_m$ is also shown in Fig. 11.

## Estimation of pO2 as a function of the degree of QE

In order to estimate an expected effect of $pO_2$ variation in tissue on $R_{QE}$ and $C_{QE}$, we assume in the extreme scenario that photons from the pick-off annihilations are separable ($R_{QE} = R_{QEsep}$), and photons from other processes (direct annihilation, pPs decay and oPs conversion) are maximally entangled ($R_{QE} = R_{QEmax}$). Thus, we assume that ($1-\alpha_m$) is equivalent to the fraction of pick-off annihilations among all 2γ annihilations, which was defined as α in equation (21):

$$1 - \alpha_m = \alpha\,. \tag{37}$$

As a consequence, involving equation (34), we obtain:

$$C_{QE} \equiv \frac{\alpha_m + 1}{2} = 1 - \frac{\alpha}{2}, \tag{38}$$

where, α reflects the dependence on $pO_2$, which may be estimated using equations (21)-(26).

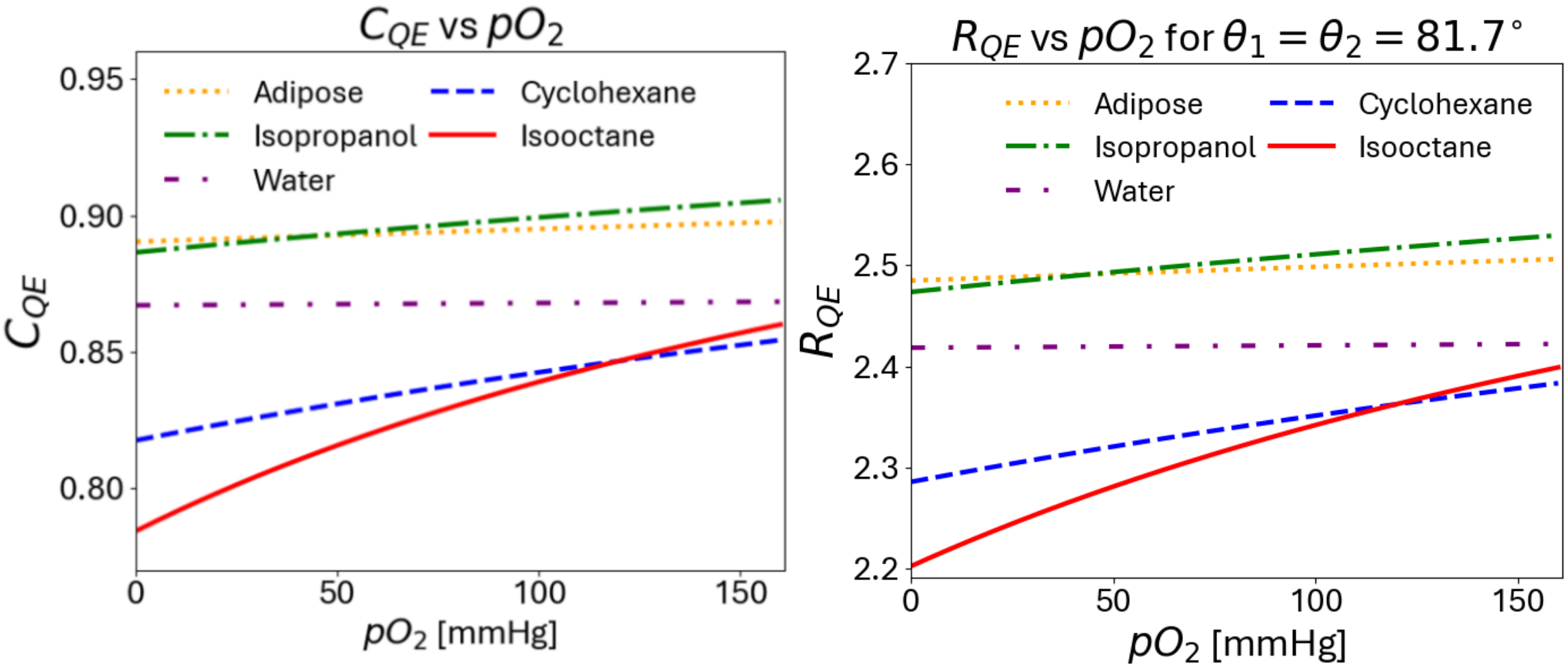

Fig. 12. Entanglement witnesses $C_{QE}$ (left) and $R_{QE}$ (right) as a function of oxygen pressure $pO_2$ estimated for water, isopropanol, cyclohexane, isooctane, and adipose. Right figure presents result of $R_{QE}$ obtained for $\theta = \theta_1 = \theta_2 = 81.7°$. Calculations were performed using earlier results for $R_{QE}$ and $C_{QE}$ as a function of α (Fig. 11), and α as a function of pO2 (Fig. 9). The calculations are performed under assumption that annihilation photons are not quantum entangled for the pick-off process and are maximally quantum entangled for other processes. This assumption may not be correct and is made only to estimate the order of magnitude for the changes of $C_{QE}$ and $R_{QE}$ as a function of oxygen pressure $pO_2$.

The assumption that $1-\alpha_m = \alpha$ is not verified experimentally, and we use it at the present stage just to get rough estimation of the order of magnitude of the effect of $pO_2$ on $R_{QE}$ and $C_{QE}$. This working assumption is based on the following observations: (i) the φ distribution for positron annihilation in metals are consistent with the predictions for maximally entangled photons [69], and in metals, positrons annihilate only directly with electrons; (ii) Decays of para-positronium, a pure quantum state, should result in maximally quantum entangled photons, this concerns also para-positronium created in the conversion of ortho-positronium on oxygen; (iii) Recent observation of non-maximal entangled photons originating from annihilation of positrons in porous polymer held in vacuum, can be explained assuming that the photons from the pick-off process are not maximally entangled (Fig. 1) [21]. The φ distribution of photons from pick-off annihilations could differ from the theoretical predictions shown by the red curves in Figs. 1 and 4B. This is because, in the case of the pick-off process, the electron and positron may annihilate while possessing a non-zero relative angular momentum.

Determination of the φ distribution separately for pick-off, direct, and para-positronium annihilations remains a challenge for future theoretical and experimental investigations. Fig. 12 shows the estimated changes in $C_{QE}$ and $R_{QE}$ as a function of the oxygen partial pressure, $pO_2$. The numerical values for the changes in $C_{QE}$ and $R_{QE}$ when $pO_2$ decreases from 50 to10 mmHg (from physoxic to hypoxic conditions) are given in Table II.

## Summary

Recent successful demonstration of positronium lifetime images of the human brain in-vivo [14], and application of positronium imaging method using clinical PET scanners [15, 17, 27-31] open perspectives for the development of in-vivo imaging of the degree of hypoxia. This is because positronium properties in the tissue depend on the partial pressure of molecular oxygen dissolved in tissue. Moreover, recent observations of the non-maximal entanglement of photons from positron-electron annihilations in matter [21] indicated that also the degree of quantum entanglement of annihilation photons may in principle inform us about the degree of hypoxia.

In this manuscript we derived formulas and estimated quantitatively how oxygen pressure affects parameters such as (i) mean ortho-positronium lifetime ($\tau_{oPs}$), (ii) the 3γ to 2γ annihilation rate ratio of ortho-positronium ($R_{oPs\text{-}3\gamma/2\gamma}$), (iii) the degree of quantum entanglement of annihilation photons expressed via entanglement witnesses $C_{QE}$ and $R_{QE}$. We also derived the formula enabling calculation of the partial pressure of oxygen based on the simultaneous measurement of $R_{oPs\text{-}3\gamma/2\gamma}$ and $\tau_{oPs}$.

We predicted the dependence of these parameters as a function of oxygen concentration for water, for organic liquids as isopropanol, cyclohexane, isooctane and for the adipose tissue. These substances were chosen to provide a baseline for approximating results in biological cells. Additionally, they represent the few substances for which all parameters required for our calculations have been previously established. We consider water because cells are mainly composed of it. However, cells also consist of molecules such as proteins, lipids, and nuclei acids. Isopropanol and isooctane are examples of organic molecules that serve as ideal solvents for lipids and hydrophobic proteins; furthermore, isopropanol, isooctane, and cyclohexane are structurally like typical cell metabolites. Adipose tissue (which is primarily composed of lipids) is currently the only tissue for which the parameters required for calculations (Table I) can be approximated.

We considered two methods for sensing hypoxia: *Method 1* utilizes decay rates of positronium, and *Method 2* utilizes the phenomenon of quantum entanglement of annihilation photons. In *Method 1*, determining hypoxia is based on simultaneous measurement of the ratio of ortho-positronium 3γ-to-2γ decay rate ($R_{oPs\text{-}3\gamma/2\gamma}$) and the mean lifetime of ortho-positronium ($\tau_{oPs}$). This method employs the fact that the ratio $R_{oPs\text{-}3\gamma/2\gamma}$ in oPs-$O_2$ process and pick-off process are different. In *Method 2*, determining hypoxia is based on the measurement of the degree of quantum entanglement of annihilation photons. This method employs the hypothesis that the degree of quantum entanglement (QE) of photons created during positronium annihilation in matter depends on the annihilation mechanism, and that the QE degree of photons from positronium annihilation in tissue is correlated with the partial pressure of oxygen dissolved in the tissue.

In order to quantitatively estimate the expected values of QE witness $C_{QE}$ and $R_{QE}$, as well as $R_{oPs\text{-}3\gamma/2\gamma}$ and $\tau_{oPs}$ as a function of the partial pressure of oxygen in tissue ($pO_2$), first we estimated the relative fraction of oPs annihilation via pick-off and via interaction with oxygen molecules in adipose, water and few organic liquids for which the data are available (Table I). Next, we derived the formula (eq. 31) that enables us to determine $pO_2$ based on the measurement of $R_{oPs\text{-}3\gamma/2\gamma}$ and $\tau_{oPs}$. Furthermore, we estimated the changes in $\tau_{oPs}$ and $R_{oPs\text{-}3\gamma/2\gamma}$ between physoxic ($pO_2$ = 50 mmHg) and hypoxic ($pO_2$ = 10 mmHg) conditions (Table II). These enabled us to determine the precision required for the measurement of $R_{oPs\text{-}3\gamma/2\gamma}$ and $\tau_{oPs}$ needed to distinguish $pO_2$ between physoxic and hypoxic conditions. In principle the precision of the $\tau_{oPs}$ measurement, $\sigma(\tau_{oPs})$, must be several times higher than these estimated changes (Table II) and in particular for adipose tissue, the precision would need to be high enough to resolve a difference in $\tau_{oPs}$ smaller than 48 ps, and difference in $R_{oPs\text{-}3\gamma/2\gamma}$ smaller than 0.0004.

As regards Method 2, for the quantitative estimation of the values of QE witness we assumed an extreme scenario that photons from pick-off process are not quantum entangled and photons from other processes are maximally quantum entangled. Under this assumption we estimated (Table II) how $C_{QE}$ and $R_{QE}$ values are changing with partial pressure of oxygen is decreasing from physoxic ($pO_2$ = 50 mmHg) to hypoxic ($pO_2$ = 10 mmHg) conditions. Based on the results obtained for adipose tissue we may estimate that in order to distinguish between physoxic and hypoxic conditions the precision would need to be good enough to see differences in $C_{QE}$ smaller than 0.0019 and difference in $R_{QE}$ smaller than 0.0055.

## Discussion

The methods of determining partial oxygen pressure based on $\tau_{oPs}$ and $R_{oPs\text{-}3\gamma/2\gamma}$ and the degree of quantum entanglement of annihilation photons may in principle be applied in the upgraded PET scanners. Here we will discuss the feasibility of the application of these methods as regards the availability of radionuclides, required statistics of registered events, and challenges due to the photons attenuation and scattering in the patient's body.

*Method 1*, which combines $\tau_{oPs\text{-}tissue}$ and $R_{oPs\text{-}3\gamma/2\gamma}$ and therefore it requires the measurement of the oPs lifetime. This is only possible when using isotopes that emit a prompt gamma in addition to a positron. The prompt gamma is needed to mark the time of oPs formation, and it is emitted e.g. by $^{44}Sc$ radionuclide ($^{44}Sc \rightarrow {}^{44}Ca^{*}\ e^{+}\ \nu \rightarrow {}^{44}Ca\ \gamma_{prompt}\ e^{+}\ \nu$, where $\nu$ denotes a neutrino which escapes from the body undetected) [13]. Hence, it will not work for PET diagnosis with fluoro-deoxy-glucose (FDG) labeled with $^{18}F$, because $^{18}F$ does not emit $\gamma_{prompt}$ ($^{18}F \rightarrow {}^{18}O\ e^{+}\ \nu$). Yet, FDG labeled with $^{18}F$ is used in 90% of all cases in PET medical practice [47].

*Method 2* for assessing hypoxia—based on measuring the degree of quantum entanglement—relies solely on annihilation photons. This method can be applied to any positron-emitting radionuclide, provided the PET scanner is capable of registering double Compton scattering within the detector material, as has already been successfully demonstrated with the J-PET scanner [21].

The sensitivity for positronium lifetime imaging of the state-of-the-art total-body PET scanners is estimated to be about 10 cps/kBq [14, 117] with first experimental values of 6.2 cps/kBq [15]. The experimental values include attenuation in the 4.5 cm diameter phantom. Assuming 6.2 cps/kBq we may expect about $10^4$ events per $cm^3$ (consistent with earlier estimations in reference [12]). This statistic is sufficient for achieving resolution of $\sigma(\tau_{oPs})$ = 10 ps organ-wise (also for large voxels as 2cm x 2cm x 2cm). Similarly, the sensitivity of the total-body PET systems for 3γ imaging is expected at the same level or even larger than these for 2γ+prompt imaging, because of the lower energies of photos from 3γ annihilations compared to photons from 2γ annihilation. Even for total-body J-PET built from plastics scintillators the sensitivity for 3γ imaging is expected at the level of about 7 cps/kBq [118]. The above considerations indicate that it can be possible to acquire tens of thousands of events organ-wise needed for the determination of $\tau_{oPs\text{-}tissue}$, $R_{oPs\text{-}3\gamma/2\gamma}$, and $C_{QE}$.

We can estimate the number of required events, knowing that the variance of the exponential lifetime spectrum, with mean lifetime equal to $\tau_{oPs}$, can be approximated by $(\tau_{oPs})^2$. For example, achieving a precision of $\sigma(\tau_{oPs})$ = 10 ps for typical values of $\tau_{oPs} \approx 2.5$ ns (Fig.10) requires the collection of 60 000 oPs events for each region of interest in the image. Similarly, achieving a precision of $\sigma(R_{oPs\text{-}3\gamma/2\gamma}) = 0.0001$ for typical values of $R_{oPs\text{-}3\gamma/2\gamma} \approx 0.022$ (Fig. 10) would also require the collection of about 50 000 oPs annihilation events. In case of isooctane with $\Delta\tau_{oPs}$ = 444 ps (Table II), $\tau_{oPs} \approx 3.5$ ns (Fig.10) and $\Delta R_{oPs\text{-}3\gamma/2\gamma} = 0.0036$ (Table II) and $R_{oPs\text{-}3\gamma/2\gamma} \approx 0.027$ (Fig.10), a statistic of 10 000 events with oPs annihilations would be sufficient. However, the attenuation of 3γ events in the body will be larger than for 2γ annihilations. A recent study indicated that the absorption in the head for $3\gamma$ is approximately 2.5 times higher than for $2\gamma$ [119].

The sensitivity for QE imaging is lower than for standard PET imaging because it requires detection of Compton scattering and subsequent interaction of scattered photons for both annihilation photons. Such double Compton scattering in coincidence constitute about 3–5% of all detected events in standard crystal PET detectors [77] but it may be larger for PET based on cadmium–zinc–telluride (CZT) [96]. While these events represent only a small fraction of the total signal, upgrading high-sensitivity total-body PET scanners, which currently reach 174 cps/kBq [121], to register them would yield an effective quantum entanglement imaging sensitivity of approximately 3 cps/kBq.

For the total-body J-PET based on plastic scintillators we expect sensitivity for QE imaging also at the level of 3 cps/kBq. Such sensitivity would enable us to collect few million events per organ using a standard PET protocol. For example, achieving a precision of e.g. $\sigma(C_{QE}) = 0.0005$ (which will be sufficient to sense a difference in $\Delta C_{QE}$ = 0.0019 expected for adipose (Table II)) for typical values of $C_{QE} \approx 0.9$ (Fig.12) a collection of about 3 million events is required for each region of interest in the image. For isooctane with $\Delta C_{QE}$ = 0.0246 (Table II) and $C_{QE} \approx 0.8$ (Fig.12), a precision of $\sigma(C_{QE})$ = 0.005 requiring 30 000 events would be sufficient.

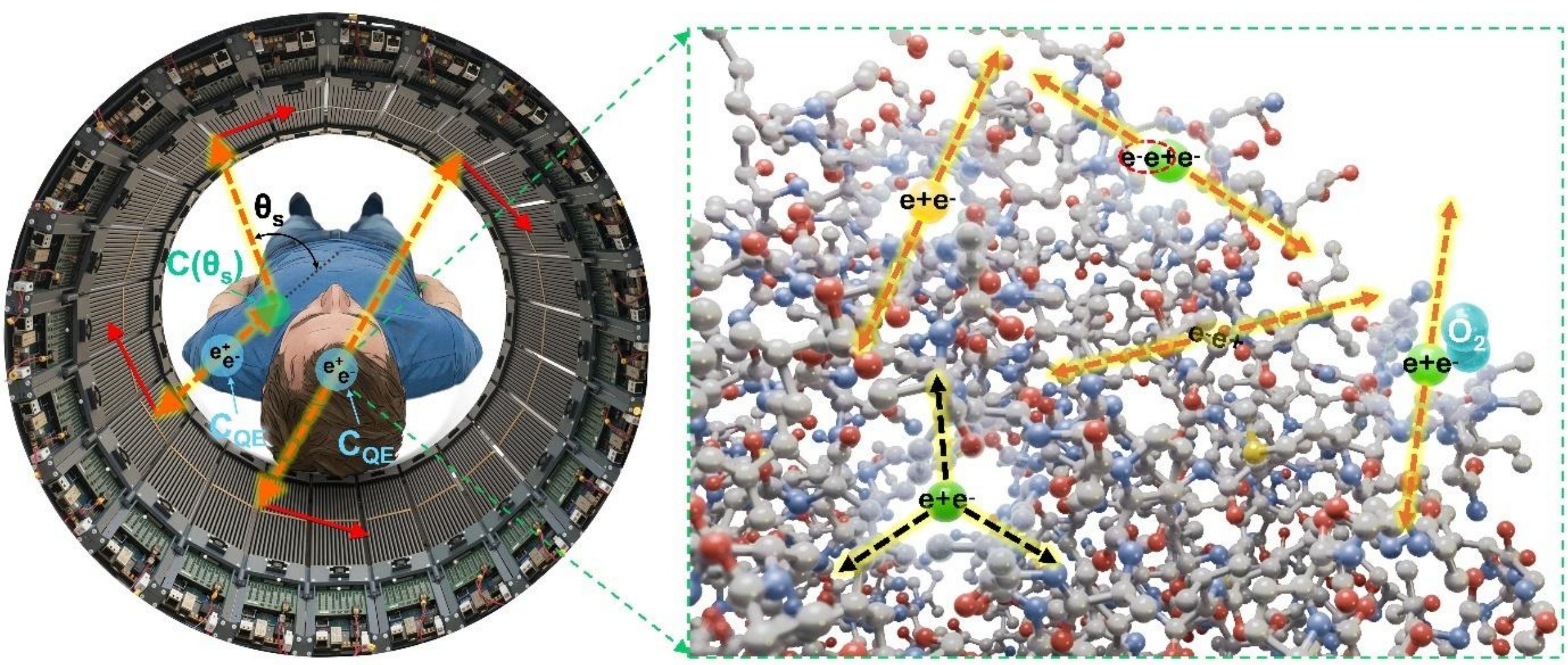

**Fig. 13.** Pictorial illustration of the measure of quantum entanglement $C(\theta_S)$ in Compton scattering in the body, and the degree of quantum entanglement $C_{QE}$ depending on the annihilation mechanism. **(Left)** Photo of the modular J-PET scanner, the first Quantum Entanglement PET, with a superimposed patient and events illustrating annihilation photons (red-yellow dashed arrows) and photons following scattering within the detector (red arrows). In one instance, the Compton scattering of an annihilation photon at an angle $\theta_S$ is indicated. **(Right)** Schematic illustration of electron-positron annihilation mechanisms, which may be characterized by different $C_{QE}$ values. The illustration uses the hemoglobin molecule as an example. Annihilation into two photons (red-yellow dashed arrows) may result from para-positronium self-annihilation (yellow circle), direct $e^+e^-$ annihilation without positronium formation, or via pick-off and conversion processes of ortho-positronium (green circles).

Finally, for the application of Quantum Entanglement (Method 2) in PET diagnostics we need to consider the effect of photons scattering inside the patient body. Typically, as much as 35% of events used for image reconstruction are distorted by at least one interaction of the annihilation photons inside the patient's body [120] (Fig. 13 left). Still few years ago it was anticipated that after the interaction of one of the photons in the body the entanglement is lost [76, 79]. Yet, unexpectedly, it was demonstrated experimentally that the φ-distribution before and after the scattering is the same [69]. Later detailed studies revealed that the entanglement degree is preserved for the forward scatterings up to about 40 degrees [70, 73, 86, 89]. Left panel of Fig. 14 shows $C(\theta_S)$ as a function of $\theta_S$ described by eq. 39. The measure of entanglement $C(\theta_S)$ as a function of the Compton scattering angle $\theta_S$ may be expressed as [70, 73]:

$$C(\Theta_S) = \frac{1 + |\cos\Theta_S| - \frac{1}{2}\sin^2\Theta_S}{1 + \cos^2\Theta_S + \frac{(1-\cos\Theta_S)^2}{2-\cos\Theta_S}}. \tag{39}$$

In PET image reconstruction, events for which photons scatter in the body at the angle $\theta_S$ larger than ~36° are already discarded by the energy requirement of limited energy loss in "energy window" applied in filtering the PET data (Fig. 14 right). This observation indicates that the residual background from photons scattered in the patient cannot be reduced using the φ-distribution. However, it indicates also that the information about tissues carried by the degree of entanglement (encoded in the $C_{QE}$ parameter (Fig. 13 right)) will not be lost if photons scatter in the body under the angle smaller than 40° [91] as indicated by vertical dashed line in Fig. 14. For the case when one annihilation photon scatters in the body (Fig. 13 left), if the degree of quantum entanglement is $C_{QE}$, then the measured value will be equal to $C_{QE} * C(\theta_S)$ as indicated in the left panel of Fig. 14 for the case of maximally entangled photons (solid curve) and for the separable photons (dashed curve).

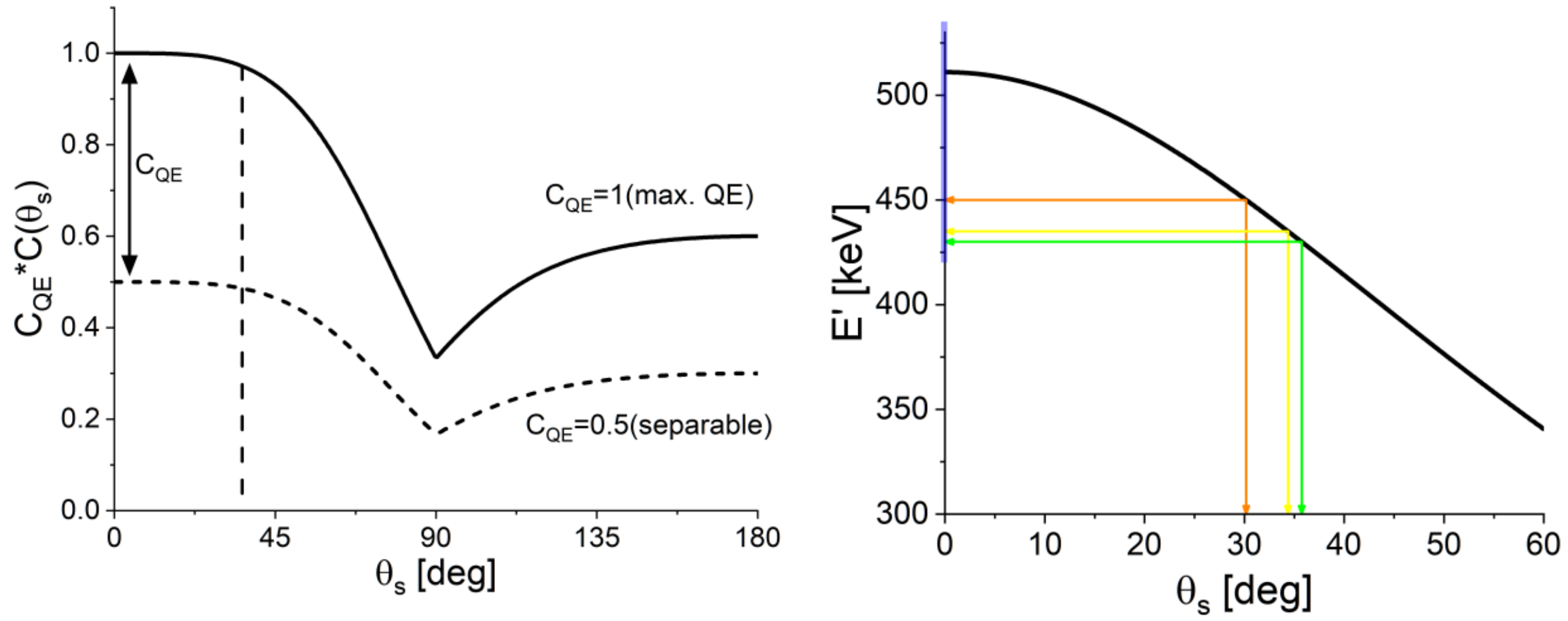

**Fig. 14.** Influence of the Compton scattering in the patient on the value of $C_{QE}$. **(Left)** The measure of entanglement of annihilation photons, $C_{QE} * C(\theta_S)$, as a function of the Compton scattering angle of one of the photons ($\theta_S$) for the maximally entangled photons (solid line), and for separable photons (dashed line) [73]. The lines show the Quantum Field Theory (QFT) predictions (eq. 39) [70, 73] for the correlation degree $C(\theta_S)$ as a function of the Compton scattering angle multiplied by the degree of entanglement $C_{QE}$, with $C_{QE}$ = 1 for maximally entangled photons, and $C_{QE}$ = ½ for separable photons [73]. The dashed vertical line indicates an angle of $\theta_S \approx 36°$ corresponding to the lower limit for angles suppressed in the state-of-the-art PET systems with the energy window starting at 430 keV. Such sharp cut is an approximation neglecting the energy resolution. **(Right)** Energy of the Compton scattered annihilation photon (E') as a function of the scattering angle $\theta_S$. The vertical axis shows the energy E' of a 511 keV annihilation photon scattered at an angle $\theta_S$. The lines indicate the lower threshold of the energy window used for scattered photon suppression in the state-of-the-art PET systems (e.g., 430–645 keV for the uExplorer [121], 430–650 keV for the uMI Panorama GS [122], 435–585 keV for the Biograph Vision Quadra [123], 450–630 keV for the PennPET Explorer [124]).

## Conclusions

This manuscript establishes a theoretical and quantitative framework for in-vivo hypoxia imaging using two novel methods based on positronium properties and quantum entanglement.

We demonstrated that $pO_2$ can be quantitatively determined using two independent approaches. *Method 1* utilizes the simultaneous measurement of the oPs mean lifetime $\tau_{oPs}$ and the 3γ-to-2γ oPs decay rate ratio $R_{oPs\text{-}3\gamma/2\gamma}$. This method relies on the distinct ratio differences between oPs-$O_2$ interactions and pick-off processes. *Method 2* introduces the degree of quantum entanglement (QE)—quantified via witnesses $C_{QE}$ and $R_{QE}$—as a novel biomarker for hypoxia, based on the correlation between the annihilation mechanism and the resulting entanglement state of the photons.

Our estimations for various media (water, organic solvents, and adipose tissue) provide a baseline for biological applications. To distinguish between physoxic ($pO_2$ = 50 mmHg) and hypoxic ($pO_2$ = 10 mmHg) conditions in adipose tissue, a precision of $\sigma(\tau_{oPs}) < 48$ ps, $\sigma(R_{oPs\text{-}3\gamma/2\gamma}) < 0.0004$ and $\sigma(C_{QE}) < 0.001$ is required. We presented arguments that such precision enabling hypoxia assessment via simultaneous imaging of $R_{oPs\text{-}3\gamma/2\gamma}$ and $\tau_{oPs}$ is achievable, at least organ-wise, when upgrading new generation of high-sensitivity total-body PET scanners [121-128] to multi-photon data acquisition presently available in J-PET [14, 129], Biograph Vision Quadra [17] and PennPET Explorer [15].

The transition of two newly postulated and introduced methods to clinical practice depends on the continuous upgrading of high-sensitivity PET scanners to support multi-photon and double Compton scattering registration. This manuscript underscores a dual role for quantum entanglement in medical diagnostics: first, as a mechanism for noise suppression in PET imaging, and more significantly, as a novel class of diagnostic biomarker.

## Acknowledgement

We are grateful for reading the earlier versions of the manuscript and constructive remarks to Prof. Shiva Abbaszadeh, Prof. Steven Bass, Dr. Neha Chug, Prof. Jerzy Dryzek, Prof. Bożena Jasińska, Mgr. Deepak Kumar, and Prof. Ewa Stepień. We also appreciate the great help in preparing the figures and calculations by Mgr. Atharva Dalvi, Mgr. Manish Das, Mgr. Deepak Kumar, Dr. Aleksander Khreptak, Dr. Bartosz Leszczyński, Mgr. Szymon Parzych, Dr. Sushil Sharma, Mgr. Keyvan Tayefi-Ardebili, and editorial corrections by Dr. Aleksander Khreptak.

We acknowledge support from the European Union within the Horizon Europe Framework Programme (ERC Advanced Grant POSITRONIUM no. 101199807), the National Science Centre of Poland through grants no. 2021/42/A/ST2/00423, 2021/43/B/ST2/02150, 2022/47/I/NZ7/03112, the Ministry of Science and Higher Education through grant no. IAL/SP/596235/2023 and SPUB/SP/627733/2025, the SciMat and qLife Priority Research Areas budget under the program Excellence Initiative – Research University at Jagiellonian University.

## Data Availability

The data used in this article are available from sources indicated in the text.

## References

[1] McKeown SR. Defining normoxia, physoxia and hypoxia in tumours-implications for treatment response. *Br J Radiol*. 2014 Mar;87(1035):20130676. doi: https://doi.org/10.1259/bjr.20130676.

[2] Swartz HM, Flood AB, Schaner PE, Halpern H, Williams BB, Pogue BW, *et al.* How best to interpret measures of levels of oxygen in tissues to make them effective clinical tools for

care of patients with cancer and other oxygen-dependent pathologies. *Physiol Rep*. 2020 Aug;8(15):e14541. doi: https://doi.org/10.14814/phy2.14541.

[3] Vaupel P, Flood AB, Swartz HM. Oxygenation status of malignant tumors vs. normal tissues: critical evaluation and updated data source based on direct measurements with $pO_2$ microsensors. *Appl Magn Reson*. 2021 Jul;52(10):1451-79. doi: https://doi.org/10.1007/s00723-021-01383-6.

[4] Cramer T, Vaupel P. Severe hypoxia is a typical characteristic of human hepatocellular carcinoma: scientific fact or fallacy? *J Hepatol*. 2022 Apr;76(4):975-80. doi: https://doi.org/10.1016/j.jhep.2021.12.028.

[5] Chen Z, Han F, Du Y, Shi H, Zhou W. Hypoxic microenvironment in cancer: molecular mechanisms and therapeutic interventions. *Signal Transduct Target Ther*. 2023 Feb 17;8(1):70. doi: https://doi.org/10.1038/s41392-023-01332-8.

[6] Gertsenshteyn I, Epel B, Giurcanu M, Barth E, Lukens J, Hall K, *et al*. Absolute oxygen-guided radiation therapy improves tumor control in three preclinical tumor models. *Front Med (Lausanne)*. 2023 Oct;10:1269689. doi: https://doi.org/10.3389/fmed.2023.1269689. Erratum in: *Front Med (Lausanne)*. 2023 Dec;10:1339872. doi: https://doi.org/10.3389/fmed.2023.1339872.

[7] Beckers C, Pruschy M, Vetrugno I. Tumor hypoxia and radiotherapy: a major driver of resistance even for novel radiotherapy modalities. *Semin Cancer Biol*. 2023 Jan;98:19-30. doi: https://doi.org/10.1016/j.semcancer.2023.11.006.

[8] Ortiz-Prado E, Dunn JF, Vasconez J, Castillo D, Viscor G. Partial pressure of oxygen in the human body: a general review. *Am J Blood Res* [Internet]. 2019 Feb 15 [cited 2026 Jan 5];9(1):1-14. Available from: https://hdl.handle.net/2445/152803.

[9] Moskal P. Positronium imaging. In: *Proceedings of the 2019 IEEE Nuclear Science Symposium and Medical Imaging Conference (NSS/MIC)*; 2019 Oct 26-Nov 2; Manchester, UK. IEEE; 2020. p. 1-3. doi: https://doi.org/10.1109/NSS/MIC42101.2019.9059856.

[10] Moskal P, Kisielewska D, Curceanu C, Czerwiński E, Dulski K, Gajos A, *et al*. Feasibility study of the positronium imaging with the J-PET tomograph. *Phys Med Biol*. 2019 Mar 7;64(5):055017. doi: https://doi.org/10.1088/1361-6560/aafe20.

[11] Moskal P, Jasińska B, Stępień EŁ, Bass SD. Positronium in medicine and biology. *Nat Rev Phys*. 2019 Jun 21;1(9):527-9. doi: https://doi.org/10.1038/s42254-019-0078-7.

[12] Moskal P, Kisielewska D, Shopa RY, Bura Z, Chhokar J, Curceanu C, *et al*. Performance assessment of the 2$\gamma$ positronium imaging with the total-body PET scanners. *EJNMMI Phys*. 2020 Jun 30;7(1):44. doi: https://doi.org/10.1186/s40658-020-00307-w.

[13] Moskal P, Dulski K, Chug N, Curceanu C, Czerwiński E, Dadgar M, *et al.* Positronium imaging with the novel multiphoton PET scanner. *Sci Adv*. 2021 Oct 13;7(42):eabh4394. doi: https://doi.org/10.1126/sciadv.abh4394.

[14] Moskal P, Baran J, Bass S, Choiński J, Chug N, Curceanu C, *et al*. Positronium image of the human brain *in vivo*. *Sci Adv*. 2024 Sep 13;10(37):eadp2840. doi: https://doi.org/10.1126/sciadv.adp2840.

[15] Huang B, Dai B, Lapi SE, Liles G, Karp JS, Qi J. High-resolution positronium lifetime tomography at clinical activity levels on the PennPET Explorer. *J Nucl Med*. 2025 Sep;66(9):1464-70. doi: https://doi.org/10.2967/jnumed.125.270130.

[16] Moskal P, Bilewicz A, Das M, Huang B, Khreptak A, Parzych S, *et al.* Positronium imaging: history, current status, and future perspectives. *IEEE Trans Radiat Plasma Med Sci*. 2025 Nov;9(8):981-1001. doi: https://doi.org/10.1109/TRPMS.2025.3583554.

[17] Mercolli L, Steinberger WM, Läppchen T, Amon M, Bregenzer C, Conti M, *et al*. *In vivo* voxel-wise positronium lifetime imaging of thyroid cancer using clinically routine I-124 PET/CT. *EANM Innov*. 2026 Mar;2:100017. doi:https://doi.org/10.1016/j.eanmi.2025.100017.

[18] Moskal P. Towards total-body modular PET for positronium and quantum entanglement imaging. In: *Proceedings of the 2018 IEEE Nuclear Science Symposium and Medical Imaging Conference (NSS/MIC)*; 2018 Nov 10-17; Sydney, NSW, Australia. IEEE; 2019. p. 1-4. doi: https://doi.org/10.1109/NSSMIC.2018.8824622.

[19] Moskal P. Positronium and quantum entanglement imaging: a new trend in positron emission tomography. In: *Proceedings of the 2021 IEEE Nuclear Science Symposium and Medical Imaging Conference (NSS/MIC)*; 2021 Oct 16-23; Piscataway, NJ, USA. IEEE; 2022. p. 1-3. doi: https://doi.org/10.1109/NSS/MIC44867.2021.9875524.

[20] Moskal P, Stępień EŁ. Positronium as a biomarker of hypoxia. *Bio-Algorithms Med Syst*. 2021 Dec 5;17(4):311-19. doi: https://doi.org/10.1515/bams-2021-0189.

[21] Moskal P, Kumar D, Sharma S, Beyene EY, Chug N, Coussat A, *et al*. Nonmaximal entanglement of photons from positron-electron annihilation demonstrated using a plastic PET scanner. *Sci Adv*. 2025 Apr 30;11(18):eads3046. doi: https://doi.org/10.1126/sciadv.ads3046.

[22] Moskal P, Stępień EŁ. Perspectives on translation of positronium imaging into clinics. *Front Phys*. 2022 Sep 16;10:969806. doi: https://doi.org/10.3389/fphy.2022.969806.

[23] Moskal P, Gajos A, Mohammed M, Chhokar J, Chug N, Curceanu C, *et al*. Testing CPT symmetry in ortho-positronium decays with positronium annihilation tomography. *Nat Commun*. 2021 Sep 27;12:5658. doi: https://doi.org/10.1038/s41467-021-25905-9.

[24] Das M, Sharma S, Beyene EY, Bilewicz A, Choinski J, Chug N, *et al*. First positronium imaging using $^{44}$Sc with the J-PET scanner: a case study on the NEMA-image quality phantom. *IEEE Transactions on Radiation and Plasma Medical Sciences*. 2026;10(4):593. doi: https://doi.org/10.1109/TRPMS.2025.3621554

[25] Das M, Sharma S, Beyene EY, Bilewicz A, Choiński J, Chug N, *et al*. First positronium lifetime imaging using $^{52}$Mn and $^{55}$Co with a plastic-based PET scanner. *Sci Rep*. Forthcoming 2026.

[26] Kubat K, Das M, Sharma S, Beyene EY, Bilewicz A, Choiński J, *et al*. First ex-vivo positronium imaging of tissues with modular J-PET scanner using $^{44}$Sc radionuclide. *Zeitschrift für Medizinische Physik*. 2026; In press. doi: https://doi.org/10.1016/j.zemedi.2026.03.004

[27] Steinberger WM, Mercolli L, Breuer J, Sari H, Parzych S, Niedzwiecki S, *et al*. Positronium lifetime validation measurements using a long-axial field-of-view positron emission tomography scanner. *EJNMMI Phys*. 2024 Aug 30;11(1):76. doi: https://doi.org/10.1186/s40658-024-00678-4.

[28] Mercolli L, Steinberger WM, Rathod N, Conti M, Moskal P, Rominger A, *et al*. Phantom imaging demonstration of positronium lifetime with a long axial field-of-view PET/CT and $^{124}$I. *EJNMMI Phys*. 2025 Aug 26;12(1):80. doi: https://doi.org/10.1186/s40658-025-00790-z.

[29] Mercolli L, Steinberger WM, Grundler PV, Moiseeva A, Braccini S, Conti M, *et al*. First positronium lifetime imaging with scandium-44 on a long axial field-of-view PET/CT. *Front Nucl Med*. 2025 Nov 20;5:1648621. doi: https://doi.org/10.3389/fnume.2025.1648621.

[30] Mercolli L, Steinberger WM, Sari H, Afshar-Oromieh A, Caobelli F, Conti M, *et al*. *In vivo* positronium lifetime measurements with a long axial field-of-view PET/CT. *medRxiv:2024.10.19.24315509* [Preprint]. 2024 Oct 22 [cited 2026 Jan 5]. doi: https://doi.org/10.1101/2024.10.19.24315509.

[31] Takyu S, Matsumoto K-i, Hirade T, Nishikido F, Akamatsu G, Tashima H, *et al*. Quantification of radicals in aqueous solution by positronium lifetime: an experiment using a clinical PET scanner. *Jpn J Appl Phys*. 2024 Aug 12;63(8):086003. doi: https://doi.org/10.35848/1347-4065/ad679a.

[32] Takyu S, Nishikido F, Tashima H, Akamatsu G, Matsumoto K-i, Takahashi M, *et al*. Positronium lifetime measurement using a clinical PET system for tumor hypoxia identification. *Nucl Instrum Methods Phys Res A*. 2024 Aug;1065:169514. doi: https://doi.org/10.1016/j.nima.2024.169514.

[33] Huang B, Wang Z, Zeng X, Goldan AH, Qi J. Fast high-resolution lifetime image reconstruction for positron lifetime tomography. *Commun Phys*. 2025 Apr 26;8:181. doi: https://doi.org/10.1038/s42005-025-02100-6.

[34] Samanta S, Sun X, Li H, Li Y. Feasibility study of positronium imaging using the NeuroEXPLORER (NX) brain PET scanner. In: *Proceedings of the 2023 IEEE Nuclear Science Symposium, Medical Imaging Conference and International Symposium on Room-Temperature Semiconductor Detectors (NSS MIC RTSD)*; 2023 Nov 4-11; Vancouver, BC, Canada. IEEE; 2023. p. 1-1. doi: https://doi.org/10.1109/NSSMICRTSD49126.2023.10338538.

[35] Shinohara N, Suzuki N, Chang T, Hyodo T. Pickoff and spin conversion of orthopositronium in oxygen. *Phys Rev A*. 2001 Sep;64(4):042702. doi: https://doi.org/10.1103/PhysRevA.64.042702.

[36] Shibuya K, Saito H, Nishikido F, Takahashi M, Yamaya T. Oxygen sensing ability of positronium atom for tumor hypoxia imaging. *Commun Phys*. 2020 Oct 1;3:173. doi: https://doi.org/10.1038/s42005-020-00440-z.

[37] Stepanov PS, Selim FA, Stepanov SV, Bokov AV, Ilyukhina OV, Duplâtre G, *et al*. Interaction of positronium with dissolved oxygen in liquids. *Phys Chem Chem Phys*. 2020 Jan 29;22:5123-31. doi: https://doi.org/10.1039/C9CP06105C.

[38] Bass SD, Mariazzi S, Moskal P, Stępień EŁ. Colloquium: positronium physics and biomedical applications. *Rev Mod Phys*. 2023 May;95(2):021002. doi: https://doi.org/10.1103/RevModPhys.95.021002.

[39] Niedźwiecki S, Białas P, Curceanu C, Czerwiński E, Dulski K, Gajos A et al. J-PET: A New Technology for the Whole-body PET Imaging. Acta Phys. Polon. B 2017;48:1567.

[40] Moskal P, Niedźwiecki S, Bednarski T, Czerwiński E, Kapłon Ł, Kubicz E, *et al*. Test of a single module of the J-PET scanner based on plastic scintillators. *Nucl Instrum Methods Phys Res A*. 2014 Nov 11;764:317-21. doi: https://doi.org/10.1016/j.nima.2014.07.052.

[41] Jasińska B, Gorgol M, Wiertel M, Zaleski R, Alfs D, Bednarski T, *et al*. Determination of the 3γ fraction from positron annihilation in mesoporous materials for symmetry violation experiment with J-PET scanner. *Acta Phys Pol B*. 2016;47(2):453-60. doi: https://doi.org/10.5506/APhysPolB.47.453.

[42] Kacperski K, Spyrou NM. Three-gamma annihilations as a new modality in PET. In: *Proceedings of the 2004 IEEE Symposium Conference Record, Nuclear Science*; 2004 Oct 16-22; Rome, Italy. IEEE; 2005. p. 3752-6. doi: https://doi.org/10.1109/NSSMIC.2004.1466696.

[43] Moskal P, Moskal I, Moskal G. TOF-PET tomograph and a method of imaging using a TOF-PET tomograph, based on a probability of production and lifetime of a positronium. Polish patent PL 227658, 2018 Jan 31; European patent EP 3039453, 2020 Apr 29; United States patent US 9851456, 2017 Dec 26.

[44] Jasińska B, Moskal P. A new PET diagnostic indicator based on the ratio of positron annihilation. *Acta Phys Pol B*. 2017;48(10):1577-82. doi: https://doi.org/10.5506/APhysPolB.48.1577.

[45] Mercurio K, Zerkel P, Laforest R, Sobotka LG, Charity RJ. The three-photon yield from $e^+$ annihilation in various fluids. *Phys Med Biol*. 2006 Aug 15;51(17):N323-9. doi: https://doi.org/10.1088/0031-9155/51/17/N05.

[46] Moskal P, Kubicz E, Grudzień G, Czerwiński E, Dulski K, Leszczyński B, *et al*. Developing a novel positronium biomarker for cardiac myxoma imaging. *EJNMMI Phys*. 2023 Mar 24;10(1):22. doi: https://doi.org/10.1186/s40658-023-00543-w.

[47] Alavi A, Werner TJ, Stępień EŁ, Moskal P. Unparalleled and revolutionary impact of PET imaging on research and day to day practice of medicine. *Bio-Algorithms Med Syst*. 2021 Nov 16;17(4):203-12. doi: https://doi.org/10.1515/bams-2021-0186.

[48] Moskal P, Stępień EŁ. Prospects and clinical perspectives of total-body PET imaging using plastic scintillators. *PET Clin*. 2020 Oct;15(4):439-52. doi: https://doi.org/10.1016/j.cpet.2020.06.009.

[49] Stepanov SV, Byakov VM. Electric field effect on positronium formation in liquids. *J Chem Phys*. 2002 Apr;116(14):6178-95. doi: https://doi.org/10.1063/1.1451244.

[50] Kotera K, Saito T, Yamanaka T. Measurement of positron lifetime to probe the mixed molecular states of liquid water. *Phys Lett A*. 2005 Sep 26;345(1-3):184-90. doi: https://doi.org/10.1016/j.physleta.2005.07.018.

[51] Ahn T, Gidley DW, Thornton AW, Wong-Foy AG, Orr BG, Kozloff KM, *et al*. Hierarchical nature of nanoscale porosity in bone revealed by positron annihilation lifetime spectroscopy. *ACS Nano*. 2021 Mar 23;15(3):4321-34. doi:https://doi.org/10.1021/acsnano.0c07478.

[52] Jean YC, Li Y, Liu G, Chen H, Zhang J, Gadzia JE. Applications of slow positrons to cancer research: search for selectivity of positron annihilation to skin cancer. *Appl Surf Sci*. 2006 Feb 28;252(9):3166-71. doi: https://doi.org/10.1016/j.apsusc.2005.08.101.

[53] Chen HM, van Horn JD, Jean YC. Applications of positron annihilation spectroscopy to life science. *Defect Diffus Forum*. 2012 Sep;331:275-93. doi: https://doi.org/10.4028/www.scientific.net/DDF.331.275.

[54] Avachat AV, Mahmoud KH, Leja AG, Xu JJ, Anastasio MA, Sivaguru M, *et al*. Ortho-positronium lifetime for soft-tissue classification. *Sci Rep*. 2024 Sep 10;14(1):21155. doi:https://doi.org/10.1038/s41598-024-71695-7.

[55] Sane P, Salonen E, Falck E, Repakova J, Tuomisto F, Holopainen JM, *et al*. Probing biomembranes with positrons. *J Phys Chem B*. 2009 Feb 19;113(7):1810-2. doi: https://doi.org/10.1021/jp809308j.

[56] Axpe E, García-Arribas AB, Mujika JI, Mérida D, Alonso A, Lopez X, *et al*. Ceramide increases free volume voids in DPPC membranes. *RSC Adv*. 2015 May;5(55):44282-90. doi: https://doi.org/10.1039/C5RA05142H.

[57] Harpen MD. Positronium: review of symmetry, conserved quantities and decay for the radiological physicist. *Med Phys*. 2004 Jan;31(1):57-61. doi: https://doi.org/10.1118/1.1630494.

[58] Jasińska B, Zgardzińska B, Chołubek G, Gorgol M, Wiktor K, Wysogląd K, *et al*. Human tissues investigation using PALS technique. *Acta Phys Pol B*. 2017;48(10):1737-1747. doi: https://doi.org/10.5506/APhysPolB.48.1737.

[59] Ore A, Powell JL. Three-photon annihilation of an electron–positron pair. *Phys Rev*. 1949 Jun;75(11):1696-9. doi: https://doi.org/10.1103/PhysRev.75.1696.

[60] Bass SD. QED and fundamental symmetries in positronium decays. *Acta Phys Pol B*. 2019;50(7):1319-33. doi: https://doi.org/10.5506/APhysPolB.50.1319.

[61] Bisi A, Fasana A, Gatti E, Zappa L. Positron annihilation in insulators. *Nuovo Cimento*. 1961 Oct;22(2):266-74. doi: https://doi.org/10.1007/BF02783017.

[62] Bussolati C, Zappa L. Anomalous three-quantum decay of positrons in alkaline earth oxides. *Phys Rev*. 1964 Nov;136(3A):A657-9. doi: https://doi.org/10.1103/PhysRev.136.A657.

[63] Al-Ramadhan AH, Gidley DW. New precision measurement of the decay rate of singlet positronium. *Phys Rev Lett*. 1994 Mar 14;72(11):1632-5. doi: https://doi.org/10.1103/PhysRevLett.72.1632.

[64] Kataoka Y, Asai S, Kobayashi T. First test of $O(\alpha^2)$ correction of the orthopositronium decay rate. *Phys Lett B*. 2009 Jan 19;671(2):219-23. doi: https://doi.org/10.1016/j.physletb.2008.12.008.

[65] Hiesmayr BC, Moskal P. Witnessing entanglement in Compton scattering processes via mutually unbiased bases. *Sci Rep*. 2019 Jun 3;9:8166. doi: https://doi.org/10.1038/s41598-019-44570-z.

[66] Acín A, Latorre JI, Pascual P. Three-party entanglement from positronium. *Phys Rev A*. 2001 Mar 19;63(4):042107. doi: https://doi.org/10.1103/PhysRevA.63.042107.

[67] Hiesmayr BC, Moskal P. Genuine multipartite entanglement in the 3-photon decay of positronium. *Sci Rep*. 2017 Nov 10;7(1):15349. doi: https://doi.org/10.1038/s41598-017-15356-y.

[68] Nowakowski M, Bedoya Fierro D. Three-photon entanglement from ortho-positronium revisited. *Acta Phys Pol B*. 2017;48(10):1955-60. doi: https://doi.org/10.5506/APhysPolB.48.1955.

[69] Ivashkin A, Abdurashitov D, Baranov A, Guber F, Morozov S, Musin S, *et al*. Testing entanglement of annihilation photons. *Sci Rep*. 2023 May 9;13(1):7559. doi: https://doi.org/10.1038/s41598-023-34767-8.

[70] Caradonna P. Kinematic analysis of multiple Compton scattering in quantum-entangled two-photon systems. *Ann Phys*. 2024 Nov;470:169779. doi: https://doi.org/10.1016/j.aop.2024.169779.

[71] Caradonna P, D'Amico I, Jenkins DG, Watts DP. Stokes-parameter representation for Compton scattering of entangled and classically correlated two-photon systems. *Phys Rev A*. 2024 Mar 20;109(3):033719. doi: https://doi.org/10.1103/PhysRevA.109.033719.

[72] Caradonna P. Compton scattering mediated by quantum entanglement. *Phys Rev A*. 2025 May 8;111(5):053708. doi: https://doi.org/10.1103/PhysRevA.111.053708.

[73] Tkachev I, Musin S, Abdurashitov D, Baranov A, Guber F, Ivashkin A, *et al*. Measuring the evolution of entanglement in Compton scattering. *Sci Rep*. 2025 Feb 19;15(1):6064. doi: https://doi.org/10.1038/s41598-025-87095-4.

[74] Žugec P, Vivoda EA, Makek M, Friščić I. A reconciliation of the Pryce-Ward and Klein-Nishina statistics for semi-classical simulations of annihilation photon correlations. *Phys. Lett. B*. 2026; 875:140346.

[75] McNamara AL, Toghyani M, Gillam JE, Wu K, Kuncic Z. Towards optimal imaging with PET: an in silico feasibility study. *Phys Med Biol*. 2014 Nov 21;59(24):7587. doi: https://doi.org/10.1088/0031-9155/59/24/7587.

[76] Toghyani M, Gillam JE, McNamara AL, Kuncic Z. Polarisation-based coincidence event discrimination: an in silico study towards a feasible scheme for Compton-PET. *Phys Med Biol*. 2016 Jul 13;61(15):5803. doi: https://doi.org/10.1088/0031-9155/61/15/5803.

[77] Kožuljević AM, Bosnar D, Kuncic Z, Makek M, Parashari S, Žugec P. Study of multi-pixel scintillator detector configurations for measuring polarized gamma radiation. *Condens Matter*. 2021 Nov 16;6(4):43. doi: https://doi.org/10.3390/condmat6040043.

[78] Caradonna P, Reutens D, Takahashi T, Takeda S, Vegh V. Probing entanglement in Compton interactions. *J Phys Commun*. 2019 Oct 11;3(10):105005. doi: https://doi.org/10.1088/2399-6528/ab45db.

[79] Watts DP, Bordes J, Brown JR, Cherlin A, Newton R, Allison J, *et al*. Photon quantum entanglement in the MeV regime and its application in PET imaging. *Nat Commun*. 2021 May 11;12(1):2646. doi: https://doi.org/10.1038/s41467-021-22907-5.

[80] Shimazoe K, Tomita H, Watts D, Moskal P, Kagawa A, Thirolf PG, *et al*. Quantum sensing for biomedical applications. In: *Proceedings of the 2021 IEEE Nuclear Science Symposium and Medical Imaging Conference (NSS/MIC)*; 2021 Oct 16-23; Piscataway, NJ, USA. IEEE; 2022. p. 1-4. doi: https://doi.org/10.1109/NSS/MIC44867.2021.9875702.

[81] Parashari S, Bokulić T, Bosnar D, Kožuljević AM, Kuncic Z, Žugec P, *et al*. Optimization of detector modules for measuring gamma-ray polarization in positron emission tomography. *Nucl Instrum Methods Phys Res A*. 2022 Oct 1;1040:167186. doi: https://doi.org/10.1016/j.nima.2022.167186.

[82] Sharma S, Kumar D, Moskal P. Decoherence puzzle in measurements of photons originating from electron-positron annihilation. *Acta Phys Pol A*. 2022 Sep;142(3):428-35. doi: https://doi.org/10.12693/APhysPolA.142.428.

[83] Shimazoe K, Uenomachi M. Multi-molecule imaging and inter-molecular imaging in nuclear medicine. *Bio-Algorithms Med-Syst*. 2022 Nov 24;18(1):127-34. doi: https://doi.org/10.2478/bioal-2022-0081.

[84] Kim D, Rachman AN, Taisei U, Uenomachi M, Shimazoe K, Takahashi H. Background reduction in PET by double Compton scattering of quantum entangled annihilation photons. *J Instrum*. 2023 Jul 4;18:P07007. doi: https://doi.org/10.1088/1748-0221/18/07/P07007.

[85] Romanchek GR, Shoop G, Abbaszadeh S. Application of quantum entanglement induced polarization for dual-positron and prompt gamma imaging. *Bio-Algorithms Med-Syst*. 2023;19(1):9-16. doi: https://doi.org/10.5604/01.3001.0054.1817.

[86] Parashari S, Bosnar D, Friščić I, Kožuljević AM, Kuncic Z, Žugec P, *et al*. Closing the door on the "puzzle of decoherence" of annihilation quanta. *Phys Lett B*. 2024 May;852:138628. doi: https://doi.org/10.1016/j.physletb.2024.138628.

[87] Kožuljević AM, Bokulić T, Grošev D, Kuncic Z, Parashari S, Pavelić L, *et al*. Investigation of the spatial resolution of PET imaging system measuring polarization-correlated Compton events. *Nucl Instrum Methods Phys Res A*. 2024 Nov;1068:169795. doi: https://doi.org/10.1016/j.nima.2024.169795.

[88] Eslami H, Mohamadian M. Optimization of the positron emission tomography image resolution by using quantum entanglement concept. *Eur Phys J Plus*. 2024 Nov 6;139:976. doi: https://doi.org/10.1140/epjp/s13360-024-05776-x.

[89] Bordes J, Brown JR, Watts DP, Bashkanov M, Gibson K, Newton R, *et al*. First detailed study of the quantum decoherence of entangled gamma photons. *Phys Rev Lett*. 2024 Sep 25;133(13):132502. doi: https://doi.org/10.1103/PhysRevLett.133.132502.

[90] Romanchek GR, Shoop G, Kupinski MA, Kuo PH, King M, Furenlid LR, *et al*. Investigation of quantum entanglement information for $\beta^+\gamma$ coincidences. *Bio-Algorithms Med-Syst*. 2024 Dec 17;20(Special Issue):27-34. doi: https://doi.org/10.5604/01.3001.0054.9079.

[91] Moskal P. Positron emission tomography could be aided by entanglement. *Physics*. 2024 Sep 25;17:138. doi: https://doi.org/10.1103/Physics.17.138.

[92] Nakamura K, Caradonna P, Fujimoto M, Uenomachi M, Shimazoe K. Monte Carlo assessment of entanglement-based positron emission tomography. *Acta Phys Pol A*. 2025; 148(6):S138. doi: 10.12693/APhysPolA.148.S138

[93] Caradonna P, Shimazoe K. Entanglement, steering, and separability in Compton-scattered annihilation photons. *Phys Rev A*. 2025 Sep 5;112(3):032413. doi: https://doi.org/10.1103/j7p7-ckp9.

[94] Shimazoe K. Prospects of nuclear medical quantum imaging in comparison to optical imaging. *JSAP Rev*. 2025;2025:250203. doi: https://doi.org/10.11470/jsaprev.250203.

[95] Kožuljević AM, Bokulić T, Grošev D, Parashari S, Pavelić L, Rade M, *et al*. Towards polarization-enhanced PET: study of random background in polarization-correlated Compton events. *Physica Medica 2026;* 145:105780. doi: https://doi.org/10.1016/j.ejmp.2026.105780

[96] Bharathi PG, Romanchek G, Shoop G, King M, Kupinski M, Furenlid L, *et al*. Machine learning-assisted event classification in cadmium zinc telluride positron emission tomography detectors leveraging entanglement-informed angular correlations. 2026; *Sci Rep* **16**:3116. doi: https://doi.org/10.1038/s41598-025-32951-6.

[97] Moskal P, Krawczyk N, Hiesmayr BC, Bała M, Curceanu C, Czerwiński E, *et al*. Feasibility studies of the polarization of photons beyond the optical wavelength regime with the J-PET detector. *Eur Phys J C*. 2018 Nov 24;78(11):970. doi: https://doi.org/10.1140/epjc/s10052-018-6461-1.

[98] Klein O, Nishina Y. Über die Streuung von Strahlung durch freie Elektronen nach der neuen relativistischen Quantendynamik von Dirac. *Z Phys*. 1929 Nov;52:853-68. doi: https://doi.org/10.1007/BF01366453.

[99] Moskal P, Alfs D, Bednarski T, Białas P, Czerwiński E, Curceanu C, *et al*. Potential of the J-PET detector for studies of discrete symmetries in decays of positronium atom – a purely leptonic system. *Acta Phys Pol B*. 2016;47(2):509-535. doi: https://doi.org/10.5506/APhysPolB.47.509.

[100] Moskal P, Czerwiński E, Raj J, Bass SD, Beyene EY, Chug N, *et al*. Discrete symmetries tested at $10^{-4}$ precision using linear polarization of photons from positronium annihilations. *Nat Commun*. 2024 Jan 2;15:78. doi: https://doi.org/10.1038/s41467-023-44340-6.

[101] Pryce MHL, Ward JC. Angular correlation effects with annihilation radiation. *Nature*. 1947 Sep 27;160:435. doi: https://doi.org/10.1038/160435a0.

[102] Kacperski K, Spyrou NM, Smith FA. Three-gamma annihilation imaging in positron emission tomography. *IEEE Trans Med Imaging*. 2004 Apr;23(4):525-9. doi: https://doi.org/10.1109/TMI.2004.824150.

[103] Kacperski K, Spyrou NM. Performance of three-photon PET imaging: Monte Carlo simulations. *Phys Med Biol*. 2005 Dec 7;50(23):5679-95. doi: https://doi.org/10.1088/0031-9155/50/23/019.

[104] Stepanov SV, Byakov VM, Stepanov PS. Positronium in biosystems and medicine: a new approach to tumor diagnostics based on correlation between oxygenation of tissues and lifetime of the positronium atom. *Phys Wave Phenom*. 2021 Jul 13;29(2):174-9. doi: https://doi.org/10.3103/S1541308X21020138.

[105] Clever HL, Battino R, Miyamoto H, Yampolski Y, Young CL. IUPAC-NIST solubility data series. 103. Oxygen and ozone in water, aqueous solutions, and organic liquids (supplement to solubility data series volume 7). *J Phys Chem Ref Data*. 2014 Aug;43(3):033102. doi: https://doi.org/10.1063/1.4883876.

[106] Kretschmer CB, Nowakowska J, Wiebe R. Solubility of oxygen and nitrogen in organic solvents from −25° to 50 °C. *Ind Eng Chem*. 1946 May;38(5):506-9. doi: https://doi.org/10.1021/ie50437a018.
[107] Battino R, Rettich TR, Tominaga T. The solubility of oxygen and ozone in liquids. *J Phys Chem Ref Data*. 1983 Apr;12(2):163-78. doi: https://doi.org/10.1063/1.555680.
[108] Weathersby PK, Homer LD. Solubility of inert gases in biological fluids and tissues: a review. *Undersea Biomed Res*. 1980 Dec;7(4):277-96.
[109] Sander R. Compilation of Henry's law constants (version 5.0.0) for water as solvent. *Atmos Chem Phys*. 2023 Oct 6;23(19):10901-12440. doi: https://doi.org/10.5194/acp-23-10901-2023.
[110] Sander R. Henry's law constants for oxygen (CAS RN 7782-44-7) [Internet]. Henry's Law Constants; [cited 2026 Jan 5]. Available from: https://henrys-law.org/henry/casrn/7782-44-7.
[111] Fellows PJ. *Food processing technology: principles and practice*. 3rd ed. Cambridge: Woodhead Publishing; 2009.
[112] Tashima H, Yamaya T. Three-gamma imaging in nuclear medicine: a review. *IEEE Trans Radiat Plasma Med Sci*. 2024;8(8):853-66. doi: https://doi.org/10.1109/TRPMS.2024.3470836.
[113] Steinberger WM, Zheng Y, Cabello J, Breuer J. Measuring 3γ ortho-positronium decay in a long-axial FOV PET/CT scanner. In: *Proceedings of the 2025 IEEE Nuclear Science Symposium (NSS), Medical Imaging Conference (MIC) and Room Temperature Semiconductor Detector Conference (RTSD)*; 2025 Nov 1-8; Yokohama, Japan. IEEE; 2025. p. 1-1. doi: https://doi.org/10.1109/NSS/MIC/RTSD57106.2025.11287847.
[114] Yu Z, Asch H, Gerber R, Redey A, Starosta K, Tam D, *et al*. Novel application of the 8π gamma-ray spectrometer for 3γ positron emission tomography. *Nucl Instrum Methods Phys Res A*. 2026 Apr;1084:171168. doi: https://doi.org/10.1016/j.nima.2025.171168.
[115] Fujimoto M, Shimazoe K, Sato R, Hamdan M, Uenomachi M, Stephenson L, *et al*. Advancing PET through direct imaging of three-photon decay using pure positron emitters. *Research Square* [Preprint]. 2025 Dec 3 [cited 2026 Jan 5]: [24 p.]. Available from: https://doi.org/10.21203/rs.3.rs-7272322/v1.
[116] Shimazoe K, Uenomachi M. Quantum sensing and imaging in the MeV regime of nuclear medicine. *Applied Physics Express*. 2026;**19(4)**:040104. doi: https://doi.org/10.1016/j.ejmp.2026.105780
[117] Parzych S, Niedźwiecki S, Beyene EY, Chug N, Conti M, Curceanu C, *et al*. Feasibility study of the positronium lifetime imaging with the Biograph Vision Quadra and J-PET tomographs. *EJNMMI Phys*. Forthcoming 2026. arXiv:2601.03172v1 [Preprint]. 2025 [cited 2026 Feb 5]: [19 p.]. Available from: https://arxiv.org/pdf/2601.03172.
[118] Ardebili KT, *et al*., in preparation
[119] Kasperska K, Skurzok M, Moskal P. Studies of attenuation effects in two- and three-photon positronium decays in phantom models. Bio-Algorithms and Med-Systems. (2025);21(Special issue (New Trends in Nuclear and Medical Physics)): 42-51. https://doi.org/10.5604/01.3001.0055.5414.
[120] Vandenberghe S, Moskal P, Karp JS. State of the art in total body PET. *EJNMMI Phys*. 2020 May 25;7(1):35. doi: https://doi.org/10.1186/s40658-020-00290-2.
[121] Spencer BA, Berg E, Schmall JP, Omidvari N, Leung EK, Abdelhafez YG, *et al*. Performance evaluation of the uEXPLORER total-body PET/CT scanner based on NEMA NU 2-2018 with additional tests to characterize PET scanners with a long axial field of view. *J Nucl Med*. 2021 Jun 1;62(6):861-70. doi: https://doi.org/10.2967/jnumed.120.250597.

[122] Li G, Ma W, Li X, Yang W, Quan Z, Ma T, et al. Performance Evaluation of the uMI Panorama PET/CT System in Accordance with the National Electrical Manufacturers Association NU 2-2018 Standard. J Nucl Med. 2024 Apr 1;65(4):652-658. doi: 10.2967/jnumed.123.265929.

[123] Prenosil GA, Sari H, Fürstner M, Afshar-Oromieh A, Shi K, Rominger A, *et al*. Performance characteristics of the Biograph Vision Quadra PET/CT system with a long axial field of view using the NEMA NU 2-2018 standard. *J Nucl Med*. 2022 Mar;63(3):476-84. doi: https://doi.org/10.2967/jnumed.121.261972.

[124] Dai B, Daube-Witherspoon ME, McDonald S, Werner ME, Parma MJ, Geagan MJ, *et al*. Performance evaluation of the PennPET Explorer with expanded axial coverage. *Phys Med Biol*. 2023 Apr 19;68(9):095007. doi: https://doi.org/10.1088/1361-6560/acc722.

[125] Godinez F, Mingels C, Bayerlein R, Mehadji B, Nardo L. Total body PET/CT: future aspects. *Semin Nucl Med*. 2025 Jan;55(1):107-15. doi: https://doi.org/10.1053/j.semnuclmed.2024.10.011.

[126] Zhang H, Ren C, Liu Y, Yan X, Liu M, Hao Z, *et al*. Performance characteristics of a new generation 148-cm axial field-of-view uMI Panorama GS PET/CT system with extended NEMA NU 2-2018 and EARL standards. *J Nucl Med*. 2024 Dec 3;65(12):1974-82. doi: https://doi.org/10.2967/jnumed.124.267963.

[127] Moskal P, Kowalski P, Shopa RY, Raczyński L, Baran J, Chug N, *et al*. Simulating NEMA characteristics of the modular total-body J-PET scanner – an economic total-body PET from plastic scintillators. *Phys Med Biol*. 2021 Aug 27;66(17):175015. doi: https://doi.org/10.1088/1361-6560/ac16bd.

[128] Ardebili KT, Beyene EY, Chug N, Curceanu C, Czerwiński E, Das M, *et al*. Development of a cost-effective total body J-PET from plastic scintillators: definitive design. In: *Proceedings of the 2025 IEEE Nuclear Science Symposium (NSS), Medical Imaging Conference (MIC) and Room Temperature Semiconductor Detector Conference (RTSD)*; 2025 Nov 1-8; Yokohama, Japan. IEEE; 2025. p. 1-2. doi: https://doi.org/10.1109/NSS/MIC/RTSD57106.2025.11287221.

[129] Korcyl G, Białas P, Curceanu C, Czerwiński E, Dulski K, Flak B, *et al*. Evaluation of single-chip, real-time tomographic data processing on FPGA SoC devices. *IEEE Trans Med Imaging*. 2018 May 17;37(11):2526-35. doi: https://doi.org/10.1109/TMI.2018.2837741.